\documentclass[nohyper,11pt,letterpaper]{JHEP3}
\usepackage{epsfig,yfonts}
\usepackage{amsfonts}
\usepackage{amsmath}

\newcommand{\beq}{\begin{equation}}
\newcommand{\eeq}{\end{equation}}
\newcommand{\beqa}{\begin{eqnarray}}
\newcommand{\eeqa}{\end{eqnarray}}
\newcommand{\beqar}{\begin{eqnarray*}}
\newcommand{\eeqar}{\end{eqnarray*}}

\newcommand{\be}{\begin{equation}}
\newcommand{\ee}{\end{equation}}
\newcommand{\bea}{\begin{eqnarray}}
\newcommand{\eea}{\end{eqnarray}}

\newcommand{\labell}[1]{\label{#1}}
\newcommand{\reef}[1]{(\ref{#1})}
\renewcommand{\eqref}[1]{(\ref{#1})}
\newcommand{\ssc}{\scriptscriptstyle}
\newcommand{\eg}{{\it e.g.,}\ }
\newcommand{\ie}{{\it i.e.,}\ }

\newcommand{\mt}[1]{\textrm{\tiny #1}}

\newcommand{\norm}[1]{\raise.3ex\hbox{:}#1\raise.3ex\hbox{:}}

\newcommand{\lsim}{\mathrel{\raisebox{-.6ex}{$\stackrel{\textstyle<}{\sim}$}}}

\newcommand{\al}{\alpha}

\newcommand{\veps}{\varepsilon}

\renewcommand{\b}{\beta}
\newcommand{\del}{\delta}

\newcommand{\ga}{\gamma}

\renewcommand{\k}{\kappa}
\renewcommand{\l}{\lambda}

\newcommand{\na}{\nabla}

\renewcommand{\t}{\theta}

\newcommand\lp{{\ell_p}}  

\newcommand\x{\times}

\def\t6 {T_\mt{D6}}
\def\t8 {T_\mt{D8}}

\def\u0{u_0}
\def\x4{{x^4}}
\def\tx4{{\tilde{x}{}^4}}
\def\hx4{{\hat{x}{}^4}}
\def\r4{{r_{\ssc \it 4}}}

\renewcommand{\c}[1]{c_\mt{#1}}
\renewcommand{\r}[1]{r_\mt{#1}}
\newcommand{\half}{\frac{1}{2}}
\newcommand{\mub}{{\bar\mu}}

\title{Holographic hydrodynamics with a chemical potential}

\author{Robert C. Myers,$^{a}$ Miguel F. Paulos$^{b}$ and Aninda Sinha$^a$ \\
$^a$ {\it Perimeter Institute for Theoretical Physics, Waterloo,
Ontario N2L 2Y5, Canada}\\
$^b$ {\it Department of Applied Mathematics and Theoretical
    Physics, Cambridge CB3 0WA, UK}\\

\vskip .5cm

{\rm E-mail:}\ \ {\tt rmyers,$\,$asinha@perimeterinstitute.ca, \
m.f.paulos@damtp.cam.ac.uk}}

\abstract{We consider five-dimensional gravity coupled to a negative
cosmological constant and a single $U(1)$ gauge field, including a
general set of four-derivative interactions. In this framework, we
construct charged planar AdS black hole solutions perturbatively and
consider the thermal and hydrodynamic properties of the plasma in
the dual CFT. In particular, we calculate the ratio of shear
viscosity to entropy density and argue that the violation of the KSS
bound is enhanced in the presence of a chemical potential. We also
compute the electrical conductivity and comment on various
conjectured bounds related to this coefficient.}


\preprint{arXiv:0903.2834 [hep-th]\\ DAMTP-2009-24}

\begin{document}{\vskip 1cm}

\section{Introduction}

Recent years have seen a remarkable confluence of string theory and
nuclear physics, with efforts to gain new theoretical insights into
the strongly coupled quark-gluon plasma (sQGP) using holographic
techniques \cite{talks}. The AdS/CFT correspondence
\cite{juan,bigRev} has proven to be a powerful tool to investigate
the thermal and hydrodynamic properties for certain strongly coupled
gauge theories \cite{review1}. Of course, the gauge theories which
are amenable to such holographic study are somewhat exotic compared
to QCD but one may suppose that certain properties of the
corresponding plasmas may be universal. The latter suggestion was
reinforced by the observation that the ratio of shear viscosity to
entropy density seemed to be a universal property of the holographic
theories yielding $\eta/s=1/4\pi$ \cite{kss1,hong}. Further,
experimental data indicates that this ratio is also unusually small
for the sQGP and even appears to yield roughly $\eta/s\sim1/4\pi$
\cite{rick}. It is now well understood that the holographic result
$\eta/s=1/4\pi$ emerges for gauge theories described by Einstein
gravity as the gravitational dual. Still this encompasses a remarkably wide class of theories and situations, \eg
with various gauge groups and matter content, with or without
chemical potentials, with non-commutative spatial directions or in
external background fields \cite{kss1,hong}. It is also well
understood that higher curvature corrections in the gravitational
dual will modify this ratio \cite{alex}--\cite{beyond}.
In fact, it was shown that for certain theories these corrections
produce even lower values \cite{cool,kats,beyond} thus
disproving{\footnote{See also \cite{mia}.}} a longstanding
conjecture that $\eta/s=1/4\pi$ represented a strict lower bound for
the viscosity of any physical system, \ie the KSS bound \cite{kss}.
Still one may interpret these new holographic calculations with
higher curvature interactions as broadening the universality class
of conformal gauge theories under study \cite{universe,beyond}.

The focus of the present paper is to use the AdS/CFT correspondence
to investigate how a nonvanishing chemical potential $\mu$ effects
the hydrodynamics of strongly coupled gauge theories. In particular,
we consider how $\eta/s$ is modified at finite $\mu$. As noted
above, with an Einstein gravity dual, the result remains
$\eta/s=1/4\pi$, even though individually the viscosity and entropy
density have a complicated dependence on $\mu$ \cite{ason,ason1}.
Hence, $\mu$ can only modify this ratio through the correction terms
appearing from higher derivative interactions in the gravitational
dual, as we will explicitly illustrate. Since the effects of the
chemical potential are tied to the higher derivative interactions,
it is interesting to examine violations of the KSS bound in this
context. For example, one might find that the chemical potential
limits any violations and that the bound is restored with
sufficiently large $\mu$, \ie $\eta/s>1/4\pi$ for $\mu>\mu_c$.
However, we identify a broad class of theories where in fact the
opposite result is found, \ie increasing $\mu$ only enhances the
violation of the KSS bound.

In principle, studying the effect of the chemical potential on
hydrodynamic properties is also of phenomenological interest. The
higher derivative modifications are associated with corrections
emerging from finite $N_c$ and $\lambda$ in the QCD plasma and these
may be significant for the sQGP \cite{qcor,universe,beyond}. So it
is again of interest to determine whether finite $\mu$ enhances or
suppresses these effects. Unfortunately the relevant chemcial
potential for baryon number is not expected to be large, \ie
$\mu_\mt{B}\sim30MeV$ or $\mu_\mt{B}/T\lsim0.15$ for recent
experiments at RHIC \cite{rhicdata} and so any effects will be
limited. However, they may still play a role as the determination of
$\eta/s$ becomes more precise in the coming years.

Turning to the holographic hydrodynamics described by the charged
black holes more broadly, we also investigate the conductivity,
$\sigma$. It was suggested that the ratio of the conductivity to the
shear viscosity could obey a bound similar to $\eta/s$ \cite{adam2}.
The heuristic reasoning behind this conjecture was as follows
\cite{adam2}: in any four-dimensional CFT, we expect $\eta \sim c
T^3$ while $\sigma \sim k T$ where $c$ and $k$ are basically central
charges of the CFT. The first of these is related to the total
number of degrees of freedom while $k$ is related to the charge
degrees of freedom. Thus it is natural to expect an upper bound on
$\sigma T^2/(\eta e^2) \propto k/c$, which in turn may be related to
the weak gravity conjecture of \cite{weak1}. While this ratio
depends on the relative normalization of the current and the stress
tensor in the CFT, it was also suggested in \cite{adam2} that this
relative normalization would not appear in the ratio of the
conductivity to the susceptibility and so it may be more natural for
this ratio to obey a universal bound. We extend this discussion to a
framework of general four-derivative interactions, as described
below.

An overview of the paper is as follows: In section \ref{eom}, we
present our action for five-dimensional gravity coupled to a
negative cosmological constant and a single $U(1)$ gauge field,
including a general set of four-derivative interactions. Further we
examine how the higher-derivative terms modify charged planar AdS
black holes, both the solution and their thermodynamic properties.
In section \ref{hydro}, we investigate the hydrodynamic properties
of these black holes with the four-derivative corrections. In
particular, we calculate both the viscosity and the conductivity of
the dual CFT plasma. Finally, a concluding discussion is
presented in section~\ref{discuss}. We also show in appendix
\ref{redef} that one can use field redefinitions to reduce the most
general four-derivative action to include only the five interactions
explicitly studied in main text.

\section{Charged black holes in higher derivative gravity} \label{eom}

We begin with five-dimensional gravity coupled to a negative
cosmological constant and a $U(1)$ gauge field in the following
action:
\begin{eqnarray}
I&=&\frac{1}{2\lp^3}\int d^5x\sqrt{-g} \left[ \frac{12}{L^2}
+R-\frac{1}{4} F^2 +\frac{\k}{3} \veps^{abcde} A_a F_{bc}F_{de}+
L^2\left(\c1 R_{abcd}R^{abcd}
 \right.\right. \labell{act1}\\
&&\qquad\qquad\quad\left.\left. +\c2 R_{abcd}F^{ab}F^{cd} +\c3
(F^2)^2 + \c4\,F^4 
+\c5\, \veps^{abcde} A_a R_{bcfg} R_{de}{}^{fg}\right)\right]\,,
 \nonumber
\end{eqnarray}
where $F^2=F_{ab}F^{ab}$ and $F^4=F^a{}_bF^b{}_cF^c{}_dF^d{}_a$. As
well as the conventional Einstein and Maxwell terms, our
two-derivative action also includes the Chern-Simons term
proportional to $\veps^{abcde}$, which naturally arises in
five-dimensional supergravity \cite{fived}. The above action
\reef{act1} also contains a general set of four-derivative
interactions. We will treat these terms in a perturbative framework
where each of the the dimensionless coefficients $\c{i}\ll1$. As
discussed in \cite{beyond}, it is natural to expect that each of
these coefficients is suppressed by a factor of $\lp^2/L^2$, which
we are assuming is very small. We demonstrate in appendix
\ref{redef} that within this perturbative framework, one can use
field redefinitions to reduce the most general four-derivative
action to include only the five interactions appearing above in
\reef{act1}.

The metric equation of motion arising from \reef{act1} is
\begin{eqnarray}
&&R_{ab}-\half R g_{ab}\ =\ \half F_{ac}F_b{}^{c}
-\frac{1}{8} F^2 g_{ab}+\frac{6}{L^2} g_{ab} \labell{geom} \\
&&\qquad+ L^2\c1 \left(\half R_{cdef}R^{cdef} g_{ab} -2
R_{(a|cde}R_{|b)}{}^{cde}+4 \nabla^c \nabla^d R_{c(ab)d}\right)
 \nonumber\\
&&\qquad+ L^2\c2\left( \half R_{cdef} F^{cd}F^{ef}g_{ab}
+3R^{cde}{}_{(a}F_{b)e} F_{cd} +2 \nabla^c \nabla^d \left( F_{c(a}
F_{b)d}\right)\right)
 \nonumber\\
&&\qquad+ L^2\c3 \left(\half (F^2)^2g_{ab} -4F^2 F_{ ac}
F_{b}{}^{c}\right) +L^2\c4\left(\half F^4
g_{ab}-4F_{ac}F^c{}_dF^d{}_eF^e{}_b  \right)
 \nonumber\\
&&\qquad+ 2L^2\c5\, \veps^{cdef}{}_{(a}\left(
R^{g}{}_{|b)ef}\nabla_g F_{cd} +2 F_{cd}\nabla_e
R_{f|b)}\right)\nonumber
\end{eqnarray}
while the vector equation of motion is given by
\begin{eqnarray}
&&\nabla_b F^{ba}+\k \veps^{abcde}F_{bc}F_{de}= -4 L^2\c2 \nabla_b
(R^{abcd} F_{cd})
 \labell{Aeom}\\
&&\qquad+8L^2 \c3 \na_b\left(F^2 F^{ba}\right) +8L^2\c4 \nabla^d
\left(F^a{}_bF^b{}_cF^c{}_d\right) - L^2\c5\, \veps^{abcde} R_{bcfg}
R_{de}{}^{fg}\,.
 \nonumber
\end{eqnarray}
As discussed above, we will solve these equations of motion
perturbatively in the coefficients $\c{i}$ of the interactions in
the four-derivative action.

\subsection{AdS/CFT dictionary}\label{diction}

With a negative cosmological constant, the gravitational theory
described by \reef{act1} naturally has an AdS$_5$ vacuum and is dual
to a four-dimensional CFT. In this holographic context, the bulk
vector field will be dual to the current generating a global $U(1)$
symmetry in the CFT. In all, the action \reef{act1} is characterized
by seven dimensionless parameters: $L^3/\lp^3$, $\kappa$ and the
five coefficients $\c{i}$. The AdS/CFT correspondence then relates
each of these gravitational couplings to various parameters that
characterize the dual field theory. For example, the holographic
framework relates the two central charges, $a$ and $c$, of
four-dimensional CFT to \cite{renorm1,central}
\begin{equation}
 \frac{L^3}{\lp^3}
\simeq\frac{c}{\pi^2}\left(1-\frac{3}{8}\frac{c-a}{c}\right)\,,\qquad
\c1\simeq\frac{1}{8}\frac{c-a}{c}\,. \labell{relate1}
\end{equation}

As described in \cite{beyond}, a key assumption in working with the
effective action \reef{act1} is that the five-dimensional gravity
theory is described by a sensible derivative expansion. That is, we
are implicitly assuming that couplings of the four- and
higher-derivative interactions are systematically suppressed by
powers of the Planck length over the (bare) AdS scale, $\lp/L$. In
particular then, we expect that $\c1\propto\lp^2/L^2\ll1$. From the
perspective of the AdS/CFT correspondence then, we are restricted by
\reef{relate1} to consider CFT's for which
 \beq
 c\sim a\gg1\qquad{\rm and}\qquad |c-a|/c\ll 1\,.
 \labell{restrict}
 \eeq
Further, our assumption about the derivative expansion in the
gravity action then restricts the size of the field theory
parameters related to the four-derivative couplings, \ie the CFT's
of interest should have the corresponding parameters being
suppressed by inverse powers of the central charge $c$.

Turning to the other couplings, it is natural to consider $\k$ and
$\c5$ together since they both appear in interactions proportional
to $\veps^{abcde}$. Both of these Chern-Simons-like terms are not
invariant under `large' gauge transformations and as a result, in
the present holographic context, they play the distinguished role of
determining anomalies for the global $U(1)$ symmetry in the dual CFT
\cite{anom,mich}. In our present analysis, we leave these
coefficients to be arbitrary constants but we should also note that
their precise values are irrelevant here since these terms play no
role below in determining the geometry or thermal properties of our
background solutions.

In contrast, the interactions parameterized by $\c2$, $\c3$ and
$\c4$ all play a role in our perturbative analysis of the charged
black holes, as well as $\c1$. From the point of view of the dual
CFT, $\c2$ characterizes the form of the three-point function of two
currents with the stress tensor \cite{hm}. Similarly, $\c3$ and
$\c4$ provide two independent couplings in the four-point function
of four currents. Again, we leave all of these coefficients
arbitrary in our general analysis. However, in the context of a
supersymmetric theory, a special case arises for the $R$-symmetry
current which is in the same supermultiplet as the stress tensor. In
this case, all of the corresponding CFT couplings will be
proportional to the difference of the central charges \cite{hm}, as
found for $\c1$ in \reef{relate1}. The holographic dual of such a
supersymetric CFT is an $N=2$ supergravity theory. While the latter
may be gauged or ungauged depending on the details of the CFT, the
dual of the $R$-symmetry current is the particular $U(1)$ vector
appearing in the five-dimensional graviton supermultiplet. In this
framework, supersymmetry dictates the form of the four-derivative
corrections to the leading supergravity action and so all of the
relevant couplings $\c{i}$ are again related \cite{mich,sugra}. This
supersymmetric setting will be of particular interest in our
discussion in section \ref{discuss}.

We should note that there is one other (dimensionless) parameter
implicit in our analysis, which can be described as the relative
normalization of the gauge and gravity kinetic terms or
alternatively as the ratio of the five-dimensional gauge coupling
to, say, the AdS scale. One can see there is an issue here since in
the conventions used in \reef{act1}, the gauge field is
dimensionless while a conventional gauge connection should have
units of energy or inverse length. Hence, we should scale the gauge
field by some appropriate scale, $A_\mu=L_*\,\tilde{A}_\mu$. With
this choice, the Maxwell term in \reef{act1} becomes
$-\frac{1}{4g_5^2}\int d^5x\sqrt{-g}\tilde{F}^2$ where the
five-dimensional gauge coupling is given by $g^2_5=2\lp^3/L_*^2$. In
any particular setting, one is typically guided by the details of
the AdS/CFT correspondence or the string theory construction to give
the proper normalization of the gauge field, \ie choosing the scale
$L_*$ --- for example, see \cite{ason,bary}. For simplicity, in the
following we make a particular convenient choice for $L_*$, but of
course it is a straightforward exercise to reinstate a general $L_*$
in our results.

To close this subsection, we observe that typically in supergravity
actions, the gauge kinetic terms will couple to various scalars.
From the dual CFT perspective, such coupling would indicate a
nontrivial three-point function mixing two currents with some scalar
operator. Hence, from this point of view, the action \reef{act1} is
not the most general since we are making a special choice for the
form of the vector kinetic term in the two-derivative action. Beyond
this choice, we note that while we are also dropping any possible
scalar couplings in the four-derivative interactions, such couplings
would only contribute at the next order in our perturbative
expansion \cite{beyond}.

\subsection{Charged black hole solutions}\label{solut}

We consider charged planar black hole solutions with the following
ansatz:\footnote{Charged black hole solutions with spherical
horizons were constructed for a general four-derivative action in
\cite{jiml}.}
\begin{eqnarray}
ds^2&=&- \frac{r^2 f(r)}{L^2}  dt^2+\frac{L^2}{r^2 g(r)}dr^2
+\frac{r^2}{L^2}(dx^2+dy^2+dz^2)\,,\labell{bkgd}\\
A_t&=& h(r)\,. \nonumber
\end{eqnarray}
The leading order solution (of the two-derivative equations of
motion) may be determined to be:
\begin{eqnarray}
f_0&=&g_0=\left(1-\frac{\r0^2}{r^2}\right)\left(1+\frac{\r0^2}{r^2}
-\frac{q^2}{\r0^2r^4}\right)\,,\labell{lead1}\\
h_0&=&\half QL^3\left(\frac{1}{\r0^2}-\frac{1}{r^2}\right)\qquad
{\rm where}\ \ Q= \frac{2\sqrt{3} q}{L^4}\,.\labell{lead2}
\end{eqnarray}
Here $\r0$ denotes the position of the (outer) event horizon. There
is also an inner horizon at
 \beq
r_-^2=\half\r0^2\left(\sqrt{1+4\frac{q^2}{\r0^6}}-1\right)\,.
\labell{inner}
 \eeq
The solution is characterized by the charge density, which is given
by $\left(*F\right)_{xyz}=Q$, and the mass density, which is
proportional to $M=\r0^4+q^2/\r0^2$. Implicity, we have fixed the
integration constant in $h_0$ such that the gauge field vanishes at
the horizon, as required by regularity.\footnote{To see this, note
that the event horizon in \reef{bkgd} is the Killing horizon where
$|\partial_t|^2 = 0$. However, as a Killing horizon, it also
contains the bifurcation surface which is a fixed point of the
Killing flow, \ie $\partial_t = 0$ on the bifurcation surface, as
opposed to the previous null condition \cite{bobw}. Hence if the
gauge field $A$ is to be a well defined one-form, then $A_t$ must
vanish there. This is, of course, the Lorentzian analog of the
topological constraint which arises more intuitively for the
corresponding Euclidean black hole. \label{foot1}} This leading
order solution is extremal with $q^2/\r0^6=2$ for which \reef{inner}
shows the two horizons coincide, \ie $r_-^2=\r0^2$. With
$q^2/\r0^6>2$, $r_-^2>\r0^2$ and the solutions actually describe the
same set of nonextremal black holes as with $q^2/\r0^6<2$ but with
$\r0$ and $r_-$ exchanging roles.\footnote{This symmetry between
$\r0$ and $r_-$ is readily seen by noting that \reef{inner} comes
from demanding the vanishing of the second factor in $f_0$, \ie
$\r0^2\, r_-^4+\r0^4\, r_-^2-q^2=0$.} Solutions where ratio of
charge to mass densities exceeds that in the extremal black hole
(\ie $Q^2/M^{3/2}>8/\sqrt{3}L^8$) are found by allowing $\r0^2$ to
become negative but, of course, such solutions all contain a naked
singularity at $r=0$.

Now we wish to construct perturbative solutions to first order in
the $\c{i}$. We maintain the ansatz \reef{bkgd} and parameterize the
perturbative solution as
\begin{eqnarray}
f(r)&=&f_0(r)(1+F(r))\,,\nonumber\\
g(r) &=& f_0(r)(1+F(r)+G(r))\,,\labell{first}\\
h(r) &=& h_0(r)+ H(r)\,,\nonumber
\end{eqnarray}
where $F(r)$, $G(r)$ and $H(r)$ are $O(\c{i})$ corrections. It is
then straightforward to solve the equations of motion \reef{geom}
and \reef{Aeom} to first order:\footnote{Our approach was as
follows: Examining the linear combination of the metric equations
\reef{geom} proportional to $G^t{}_t-f/g\,G^r{}_r$ (where $G^a{}_b$
is the Einstein tensor), one finds a first-order linear ODE for
$G(r)$ which is readily soluble. Given $G(r)$, the $t$ component of
the vector equations \reef{Aeom} is easily solved for $H(r)$.
Finally with the solution for these two perturbations, $F(r)$ can be
determined by solving the first-order linear ODE coming from the
$G^r{}_r$ equation.}
\begin{eqnarray}
6 f_0 F(r) &=& 2 \left(2 c_1-3
g_1\right)+\frac{\r0^4}{r^4}f_1+12\frac{q^2}{r^6}\left(h_2-26\c1-12\c2\right)
+12\frac{\r0^8}{r^8}\left(1+\frac{q^2}{\r0^6}\right)^2\c1
\labell{ff} \\
&&\qquad+8\frac{q^2\r0^4}{r^{10}}\left(1+\frac{q^2}{\r0^6}\right)(5\c1+6\c2)
+\frac{q^4}{r^{12}}\left(17\c1-24(\c2+6\c3+3\c4)\right)\,,\nonumber\\
G(r)&=& g_1-\frac{8}{3}\frac{q^2}{r^6}(13 \c1+12 \c2)\,,\labell{gg}\\
H(r) &=& h_1-\sqrt{3}\frac{q}{L r^2}h_2-8 \sqrt{3} \frac{q\r0^4}{L
r^6}\left(1+\frac{q^2}{\r0^6}\right)\c2
\nonumber \\
&&\qquad + \frac{1}{\sqrt{3}}\frac{q^3}{ L r^8}(48 \c2+144 \c3+72
\c4-13 \c1)\,,\labell{hh}
\end{eqnarray}
where $f_1,$ $g_1$, $h_1$ and $h_2$ are (dimensionless) integration
constants.

We fix the integration constants as follows:
\begin{itemize}
\item
The background metric for the dual CFT can be extracted from the
asymptotic behaviour in the black hole metric \reef{bkgd} as
 \beq
ds^2_\mt{CFT} =-f_\infty dt^2 + dx^2+dy^2+dz^2 \labell{cftmet}
 \eeq
where we defined $f_\infty\equiv f(r\rightarrow\infty)$. Hence to
fix the speed of light to be one in the dual gauge theory, we
require that $f_\infty=1$. From \reef{ff}, we find
 \beq
g_1=\frac{2}{3}\c1\,. \labell{gg1}
 \eeq
Note that this fixes the asymptotic behaviour
$g(r\rightarrow\infty)\rightarrow 1 + 2\c1/3$ which reflects the
fact that, as noted in \cite{beyond}, the AdS scale of the
background geometry is perturbed to be
 \beq
\frac{1}{\hat{L}^2}=\frac{1}{L^2}\left(1+\frac{2}{3}\c1\right)\,.
\labell{shiftL}
 \eeq
when $\c1$ is nonvanishing.
\item
For simplicity, we require a regular $F(r)$ and fix the position of the event horizon to remain at $r=\r0$. That
is, we require that $f_0F(r=\r0)=0$. Again from \reef{ff}, this
fixes $f_1$ to be
 \beqa
f_1&=& -2 \left(8 c_1-3
g_1\right)+4\frac{q^2}{\r0^6}\left(62\c1+24\c2-3h_2\right)
\labell{ff1} \\
&&\qquad\qquad\quad
-\left(\frac{q^2}{\r0^6}\right)^2\left(69\c1+24(\c2-6\c3-3\c4)\right)\,.
\nonumber
 \eeqa
\item
We require that $A_t$ vanishes on the horizon --- as
described in footnote \ref{foot1}. Setting $H(r=\r0)=0$ in
\reef{hh}, we fix $h_1$ to be
 \beq
 h_1=\sqrt{3}\frac{q}{L \r0^2}(h_2+8\c2)
+\frac{1}{\sqrt{3}}\frac{q^3}{ L \r0^8}(13 \c1-24 (\c2+6 \c3+3
\c4))\,.
 \labell{hh1}
 \eeq
\item
Finally we may use the remaining freedom to require that the charge
density is fixed as in the leading order solution, \ie
$\left(*F\right)_{xyz}=Q$. The perturbed vector equation of motion
\reef{Aeom} can be written in the form $\nabla_bX^{ba}=0$ for the
appropriate antisymmetric tensor $X_{ab}$. In the perturbed
solution, $\left(*X\right)_{xyz}$ is a constant independent of
radius and it is natural to define the charge density to be
$\left(*X\right)_{xyz}=Q$. This allows us to fix $h_2$, which is
most simply done by examining this constraint for asymptotic $r$
where
 \bea
\lim_{r\rightarrow\infty}\left(*X\right)_{xyz}&=&
\lim_{r\rightarrow\infty}\left[\frac{r^3}{L^3}\sqrt{g/f}\left(F_{rt}-8L^2\c2\,
R_{rt}{}^{rt}F_{rt}\right)\right]
 \nonumber\\
 &\simeq&\left(1+\frac{1}{2}g_1+h_2+8\c2\right)Q
\,.\labell{newQ}
 \eea
Hence we fix
 \beq
h_2=-\frac{1}{2}g_1-8\c2=-\frac{1}{3}\c1-8\c2\,, \labell{AA2}
 \eeq
where in the last expression, we substituted for $g_1$ as in
\reef{gg1}.
\end{itemize}

\subsection{Black hole thermodynamics}\label{thermo}

For the thermodynamics of the above charged black holes, let us
begin by first reviewing the results for the leading order solution,
\reef{lead1} and \reef{lead2}.\footnote{The thermodynamics of
charged AdS black holes has been well studied \cite{charge,charge1},
of course, but the focus was on solutions with spherical horizons.}
The temperature of the dual CFT is precisely the Hawking temperature
calculated as the inverse of the periodicity of time in the
corresponding Euclidean solution:
 \be
T=\frac{\r0}{\pi L^2} \left(1-\frac{q^2}{2
\r0^6}\right)\,.\labell{temp0}
 \ee
Note that the temperature vanishes for the extremal black hole with
$q^2/\r0^6=2$.

Next we would like to consider the chemical potential of the system
which is related to the asymptotic value of the potential $A_t$.
However, as discussed in section \ref{diction}, we must scale the
gauge field by some appropriate scale, $A_\mu=L_*\,\tilde{A}_\mu$,
to produce a chemical potential with the appropriate units of
energy. The chemical potential then becomes
 \beq
\mu=\lim_{r\rightarrow\infty}\tilde{A}_t=\frac{\sqrt{3}\,q}{L_*L\,\r0^2}.
 \labell{chem0}
 \eeq
In any particular setting, one would be guided by the details of the
AdS/CFT construction or the string theory construction to give the
proper normalization of the chemical potential. However, for
simplicity in our general analysis, we will make the convenient
choice
 \be
L_*=\pi \,L\labell{choice}
 \ee
in the following. Of course, it is a straightforward exercise to
reinstate a general $L_*$ in the following calculations. Note that
with the preceding choice, we may write
 \be
r_0=\pi L^2 \frac{T}{2}\left(1+\sqrt{1+\frac{2}{3}\frac{\mu^2}{T^2}}
\right)\,. \labell{rad0}
 \ee
Further in the extremal limit $T=0$, the horizon radius remains
finite with $r_0=\pi L^2\,\mu/\sqrt{6}$. It will also be convenient
to denote the ratio of the chemical potential to the temperature as
 \be
\mub\equiv\frac{L^*}{\pi L}\,\frac{\mu}{T}=
\frac{\sqrt{3}\frac{q}{\r0^3}}{1-\frac{q^2}{2 \r0^6}}\,.
 \labell{defmu}
 \ee
This formula may be inverted to yield
 \be
\frac{q}{\r0^3}=\frac{2}{\sqrt{3}}\mub\,\left(1+\sqrt{1+\frac{2}{3}\mub^2}
\right)^{-1}\simeq\frac{\mub}{\sqrt{3}}\left(1-\frac{1}{6}\mub^2+
\frac{1}{18}\mub^4+\cdots\right)\,, \labell{express0}
 \ee
where the last expression is a Taylor series for small
$\mub$.\footnote{We provide such a Taylor series expansion for all
of our results with an eye towards the fact that $\mu_\mt{B}/T$ is
small at RHIC.}

To proceed further, we apply the standard path integral techniques
\cite{hawk} in which we identify the Euclidean action $I_E = W/T$,
where $W(T, \mu)$ is the Gibbs free energy, \ie the thermodynamic
potential in the grand canonical ensemble.\footnote{One could also
consider the microcanonical ensemble with a fixed charge density
$n_q$ by making the standard Legendre transform to the Helmholtz
free energy: $F(T,n_q)=W(T,\mu)+\int d^3\!x\, n_q\, \mu$. In the
AdS$_5$ language, this corresponds to adding an additional boundary
term to the action which ensures that the appropriate boundary
condition corresponds to fixing the (radial) electric field, rather
than the gauge potential, at asymptotic infinity
--- for example, see \cite{bary,charge}.}
To calculate the Euclidean action, as well as the bulk action
\reef{act1}, one includes the Gibbons-Hawking boundary term
\cite{hawk} and the appropriate boundary `counter-term' action
\cite{renorm1,renorm2}. Alternatively, one can use background
subtraction and consider the difference in the (bulk) action for the
charged black hole and for AdS$_5$ with a constant gauge potential.
Further since we are considering planar black holes \reef{bkgd}, the
spatial volume in the dual CFT is infinite and so we divide our
result for any extensive quantities by a regulator volume $V_x$ and
work with the corresponding density. We do not go into the details
of the calculations here but only present the final result for the
free energy density:
 \beqa
w&=&-\frac{\r0^4}{2\lp^3L^5}\left(1+\frac{q^2}{\r0^6}\right)
 \nonumber\\
&=&-\frac{1}{2\lp^3L^5}\left(\r0^4+\frac{\pi^2L^4}{3}\mu^2\r0^2
\right)\,.\labell{free}
 \eeqa
While in principle we could use \reef{rad0} to express the free
energy density entirely in terms of $T$ and $\mu$, the last
expression above with $w\left(\r0(T,\mu),\mu\right)$ is sufficient
for most calculations. In particular, the standard thermodynamic
identities yield the entropy density and the charge density:
 \beqa
s&=&-\left.\frac{\partial w}{\partial
T}\right|_{\mu}=\frac{2\pi\,\r0^3}{\lp^3L^3}\,,
 \labell{ent0}\\
n_q&=&-\left.\frac{\partial w}{\partial \mu}\right|_{T}=
\frac{\sqrt{3}\pi}{\lp^3L^3} q\,.
 \labell{char0}
 \eeqa
Note that the result for the entropy density matches the expected
result for the Bekenstein-Hawking entropy of the black hole horizon.
Using the previous expressions, we can also express these quantities
in terms of the temperature and chemical potential:
 \beqa
s&=&\frac{\pi^4L^3}{4\lp^3} T^3 \left(1+
\sqrt{1+\frac{2}{3}\mub^2}\,\right)^3
 \nonumber\\
&\simeq&\frac{2\pi^4L^3}{\lp^3} T^3
\left(1+\frac{1}{2}\mub^2+\frac{1}{216}\mub^6+\cdots\right)\,,
 \labell{entro00}\\
n_q&=& \frac{\pi^4L^3}{4\lp^3} \mu\,
T^2\left(1+\sqrt{1+\frac{2}{3}\mub^2}\,\right)^2
 \nonumber\\
&\simeq&\frac{\pi^4L^3}{\lp^3} \mu\,
T^2\left(1+\frac{1}{3}\mub^2-\frac{1}{36}\mub^4+
\frac{1}{108}\mub^6+\cdots\right)\,,
 \labell{char00}
 \eeqa
where in both cases, we have also given a Taylor series for small
$\mub$. Hence for large $T$ (or small $\mu/T$) our holographic model
yields $s\propto T^3$ and $n_q\propto\mu\, T^2$, as may have been
anticipated. We can also combine the above expressions, \reef{ent0}
and \reef{char0}, to calculate the energy density:
 \beqa
\rho_E&=&w+T\,\!s+\mu\,n_q \nonumber\\
&=&\frac{3}{2}\frac{\r0^4}{\lp^3L^5}\left(1+\frac{q^2}{\r0^6}\right)
=\frac{3}{2}\frac{M}{\lp^3L^5}\,. \labell{ener0}
 \eeqa

We now turn to the first order solution \reef{first} which takes
into account $O(\c{i})$ terms in the equations of motion. The temperature of
the perturbed black hole becomes
 \bea
T&=&\frac{r_0}{\pi L^2} (1-\frac{q^2}{2 r_0^6})\left(1+
F(r_0)+\frac{1}{2} G(r_0)\right)
 \labell{temp1}\\
&=&\frac{r_0}{\pi
L^2}\left[1-\frac{5}{3}\c1-\frac{q^2}{2\r0^6}\left(
1+\frac{31}{3}\c1+16\c2\right)-\left(\frac{q^2}{\r0^6}\right)^2
\left(9\c1-4\c2-24\left(2\c3+\c4\right)\right)\right] \,.
 \nonumber
 \eea
Note that the corrections shift the condition for the extremal limit
$T=0$ to be
 \be
\frac{q^2}{\r0^6}=2\left[ 1-48 \left(\c1-2(2\c3+\c4)\right) \right] \,.\labell{extrem1}
 \ee
The asymptotic value of $A_t$ determines the chemical potential, as
in \reef{chem0}. The modified result is given by
 \bea
\mu&=& \frac{\sqrt{3}\,q}{\pi L^2r_0^2}+ \frac{h_1}{\pi L}
 \labell{chem1}\\
&=&\frac{\sqrt{3}\,q}{\pi L^2r_0^2}\left[1-\frac{1}{3}\c1+8\c2
+\frac{q^2}{r_0^6}\left(\frac{13}{3}\c1-8\c2-24\left(2\c3+\c4
\right)\right)\right]\,,
 \nonumber
 \eea
where we have chosen $L_*$ as in \reef{choice} and $h_1$ is fixed as
in \reef{hh1}.

The free energy density is most easily calculated using background
subtraction, as described above, with the final result:
 \beqa\labell{free1}
w&=&-\frac{\r0^4}{2\lp^3L^5}\left[1+\frac{19}{3}\c1+\frac{q^2}{\r0^6}
\left(1-\frac{113}{3}\c1-32\c2\right)
+\frac{q^4}{\r0^{12}}\left(\frac{23}{2}\c1+4\c2-12(2\c3+\c4)\right)
\right]
 \nonumber \\
&=& -\frac{\pi^4 L^3}{2 \lp^3}T^4\left[1+13 \c1+(1+{11\over
3}\c1)\mub^2  \right. \\ &&
\qquad\qquad\qquad\qquad+\left.\left({1}+\frac{26}{3}
\c1+24(\c2+2\c3+\c4)\right)\frac{\mub^4}{6}
-\frac{1}{108}(1-15\c1)\mub^6+\cdots\right]\,.\nonumber
 \eeqa
Now using the same thermodynamic identities as above, we arrive at
the following expressions for the entropy and charge densities:
 \bea
s&=&\frac{2 \pi  r_0^3}{\lp^3L^3}\left(1+8 \c1 -4 (7 \c1+6 \c2)
\frac{q^2 }{r_0^6}\right)\,,
 \nonumber\\
 &=&\frac{\pi^4L^3}{4\lp^3} T^3 \left[\left(1+
\sqrt{1+\frac{2}{3}\mub^2}\,\right)^3 +\frac{\c1}{3} \left(1+
\sqrt{1+\frac{2}{3}\mub^2}\,\right)\left(78-2\mub^2
+6\frac{13+4\mub^2}{\sqrt{1+\frac{2}{3}\mub^2}} \right)\right]
 \nonumber\\
&\simeq&\frac{2\pi^4L^3}{\lp^3} T^3
\left[1+13\c1+\frac{\mub^2}{2}\left(1+\frac{11}{3}\c1\right)
+\frac{\mub^6}{216}\left(1-15\c1\right)+\cdots\right]\,,
 \labell{ent1}\\
n_q &=& \frac{\pi^4L^3}{4\lp^3} \mu\, T^2
\left[\left(1+\sqrt{1+\frac{2}{3}\mub^2}\,\right)^2
+\frac{\c1}{3}\left(33+46\mub^2
+\frac{33-4\mub^2}{\sqrt{1+\frac{2}{3}\mub^2}}\right)
+32\left(\c2+2\c3+\c4\right)\mub^2\right]
 \nonumber\\
&\simeq&\frac{\pi^4L^3}{\lp^3} \mu\,
T^2\left[1+\frac{11}{3}\c1+\frac{\mub^2}{3}
\left(1+\frac{26}{3}\c1+24(\c2 +2\c3+\c4)\right)\right.
 \nonumber\\
&&\qquad\qquad\qquad\qquad\qquad \left.
-\frac{\mub^4}{36}\left(1-15\c1\right)+
\frac{\mub^6}{108}\left(1-\frac{73}{3} \c1\right)+\cdots\right]\,,
 \labell{char1}
 \eea
Note that the first expression for the entropy density matches
precisely the result found using Wald's formula for higher curvature
theories \cite{walds}. It is interesting to observe that when the
entropy is expressed in terms of $T$ and $\mu$, it becomes
independent of $\c2$, which appears in the original `geometric'
expression for the entropy. In contrast, all of the coefficients,
$\c1$, $\c2$, $\c3$ and $\c4$, appear in the charge density. We also
note here that just as for the leading order results, one finds
$\rho_E=-3 w$ when the first order corrections are included.


\section{Holographic hydrodynamics} \label{hydro}

We now turn to computing the shear viscosity and the conductivity of
the holographic plasma represented by the charged black holes in the
previous section. We follow the approach of expressing the transport
coefficients in terms of field theory correlators using the Kubo
formula and then calculating these correlators with holographic
techniques \cite{hydro}. However, as we are working with the higher
curvature interactions in \reef{act1}, we must generalize these
standard calculations. The analogous calculations for a particular
four-curvature interaction first appeared in \cite{alex} and the
latter are readily adapted to the present case. Our presentation
also builds on the recent work of \cite{hong} which focussed on the
hydrodynamic limit of low frequency and momenta, showing that there
is a simple relation between quantities computed in the membrane
paradigm approach and those calculated using AdS/CFT. In particular,
the shear viscosity of a field theory with a gravity dual may be
related in a simple fashion to the membrane coupling constant of a
certain minimally coupled scalar in the dual gravitational
background. Our calculation closely parallels the discussion of
\cite{naba} which presented a general framework to calculate the
shear viscosity for higher curvature theories.

Before proceeding, we make a change of coordinates $u=r_0^2/r^2$
which is more readily adapted to the hydrodynamic calculations. With
this coordinate choice, the background solution \reef{bkgd} becomes
\begin{eqnarray}
ds^2&=&- \frac{\r0^2 }{L^2}\frac{f(u)}{u}  dt^2+\frac{L^2}{4 u^2
g(u)}du^2
+\frac{\r0^2 }{L^2}\frac{1}{u}(dx^2+dy^2+dz^2)\,,\labell{bkgd1}\\
A_t&=& h(u)\,, \nonumber
\end{eqnarray}
where the leading order solution \reef{lead1} and \reef{lead2} now takes
the form:
\begin{eqnarray}
f_0&=&g_0=\left(1-u\right)\left(1+u
-\frac{q^2}{\r0^6}u^2\right)\,,\labell{lead3}\\
h_0&=&\frac{\sqrt{3}q}{L\r0^2}\left(1-u\right)\,.\labell{lead4}
\end{eqnarray}
Of course, one must also make the appropriate substitution in the
perturbative solution given by \reef{first}--\reef{hh}. A
simplifying feature, however, is that with our choice of the
integration constants described above, in particular for $f_1$, the
event horizon remains fixed at $r=\r0$ or $u=u_0=1$ in the
perturbative solution. Of course, in terms of the new radial
coordinate, the asymptotic boundary corresponds to $u=0$.

\subsection{Corrections to $\displaystyle {\eta/s}$}
\label{shear}

Kubo's formula relates the shear viscosity to the low frequency and
zero momentum limit of the retarded Green's function of the stress
tensor in the CFT
 \be
G^R_{xy,xy}(\omega,\mathbf k=0)=-i \int dt d{\mathbf x}\,e^{i\omega t}
\theta(t)\,\langle [T_{xy}(x), T_{xy}(0)]\rangle\,.\labell{kubo}
 \ee
Concretely one has
 \be
\eta=-\lim_{\omega\to 0} \frac 1{\omega}\, \mbox{Im}~
G^R_{xy,xy}(\omega,\mathbf k=0)\,.\labell{Kubo}
 \ee
The retarded Green's function may be computed using the prescription
first set out in \cite{hydro}. Translating the calculation of the
correlator to a holographic one, one first finds the effective
action for the metric perturbation $h_x{}^y(t,u)=\int
\frac{d^4k}{(2\pi)^4}\, \phi_k(u)e^{-i \omega t+ik z}$. Evaluating
the action \reef{act1} to quadratic order in the fluctuations
$\phi_k(u)$ yields
 \bea
 I_{\phi}^{(2)}&=& \frac 1{2\lp^3} \int \frac{d^4k}{(2\pi)^4} du
\left(A(u) \phi''_k \phi_{-k}+B(u) \phi'_k \phi'_{-k}+C(u) \phi'_{k}
\phi_{-k}
 \right.\nonumber \\
&&\left.\qquad\qquad+D(u) \phi_k \phi_{-k} +E(u) \phi''_k
\phi''_{-k}+F \phi''_k \phi'_{-k}\right) +\mathcal K
\label{EfAction}. \eea
This form of the effective action originally appeared in
\cite{alex}, where the effect of certain $R^4$ terms were
considered, but this general form will arise for any action
involving any powers of the curvature tensor (but not derivatives of
the curvature\footnote{A generalization to include derivatives of
the curvature appears in \cite{naba}.}) and so appears again in the
present context with the action \reef{act1}. If we consider an
action where the higher derivative terms come coupled through some
parameter $\gamma$, then the two functions $E$ and $F$ are $
O(\gamma)$. Following \cite{alex},  we have also added a generalized
Gibbons-Hawking boundary term $\mathcal K$ in \reef{EfAction},
 \be
\mathcal K=\frac 1{2\lp^3} \int \frac{d^4k}{(2\pi)^4}\left.
\left(K_1+K_2+K_3\right)\right|^{u=1}_{u=0}\,.\labell{gibhawk}
 \ee
with
 \bea
K_1&=&-A \phi'_k \phi_{-k}\, \qquad\quad 
K_2=-\frac F2 \phi_k' \phi'_{-k} \labell{gibhawk2}\\
K_3&=&E\left(p_1\, \phi'_k+2 p_0\, \phi_k\right)\phi'_{-k}
 \nonumber
 \eea
The first term $K_1$ is essentially the contribution of original
Gibbons-Hawking term while $K_2$ and $K_3$ are new $O(\gamma)$
contributions. In the term $K_3$, the coefficient functions
$p_0,p_1$ are defined in terms of the linearized equation of motion
for $\phi$:
 \be
\phi''+p_1\, \phi'+p_0\,\phi=\mathcal O(\gamma)\,.
 \labell{lineom}
 \eeq
With this boundary term \reef{gibhawk}, the variational principle is
valid up to $\mathcal O(\gamma^2)$ \cite{alex}.

Given the effective action, let us proceed in trying to compute the
shear viscosity. First, we integrate by parts to rewrite the action
in a more symmetric fashion:
 \bea
\tilde I_{\phi}^{(2)}&=&\frac 1{2\lp^3} \int \frac{d^4k}{(2\pi)^4}
du \left((B-A-F'/2) \phi'_k \phi'_{-k}+E \phi''_k \phi''_{-k}\right.
 \nonumber \\
&& \qquad\qquad\left.+(D-(C-A')'/2)\phi_k
\phi_{-k}\right)+\tilde{\mathcal K}\,.
 \labell{EfAction2} \eea
The integration by parts eliminates $K_1$ and $K_2$ in the boundary
term and $\tilde{\mathcal K}$ is given by:
\be \tilde{\mathcal K}=\frac 1{2l_p^3} \int
\frac{d^4k}{(2\pi)^4}\left.\left(K_3+\frac 12 (C-A') \phi_k
\phi_{-k})\right) \right|^{u=1}_{u=0} \ee

At this point, it is convenient to define the (radial) canonical
momentum for our effective scalar as:
 \be
\Pi_k(u)\equiv\frac{\delta \tilde I_{\phi}^{(2)}}{\delta
\phi_{-k}'}=\frac{1}{\lp^3}\left((B-A-F'/2) \phi_k'(u)-(E
\phi_k''(u))'\right)\,,
 \labell{CanM}
 \ee
where in the variation we regard
$\phi''$ as $(\phi')'$. The scalar equation of motion then takes the
simple form
 \be
\partial_u \Pi_k(u)=M(u)\, \phi_k(u)\,,
\qquad M(u)\equiv \frac{1}{\lp^3}(D-(C-A')'/2)\,.
 \labell{lineom2}
 \ee
together with the definition of the canonical momentum \reef{CanM}.

To compute the retarded Green's function, we now evaluate the
effective action on-shell, which reduces to a boundary term using the linearized equation of motion \reef{lineom2}
 \be
I_{\mt{on-shell}}= \int \frac{d^4k}{(2\pi)^4}\ \mathcal F_k
\big|_{u=0}^{u=1} \labell{onshell}.
 \ee
The retarded Green's function is then given by the flux factor
\cite{hydro} evaluated at the asymptotic boundary
 \be
G^R_{xy,xy}(\omega,\mathbf k)= -\lim_{u\to 0} \frac{2\,\mathcal
F_k}{\phi_k(u) \phi_{-k}(u)}
 \labell{green}
 \ee
where the factors $\phi_k(u) \phi_{-k}(u)$ ensure the appropriate
normalization for the Green's function. Implicitly, this expression
is also evaluated on $\phi(u)$ with infalling boundary conditions at
the horizon. In the present case, the flux factor reduces to%
 \bea
2\mathcal F_k &=&  \Pi_k \phi_{-k}+(C-A')\phi_k \phi_{-k}+E \phi_k''
\phi_{-k}'+K_3
 \nonumber\\
 &=& \Pi_k \phi_{-k}+(C-A')\phi_k \phi_{-k}+E p_0 \phi'_k \phi_{-k}
 \labell{flux},
  \eea
where on the second line we have used the lowest order equation of
motion \reef{lineom} for $\phi_k$. We see that the flux is given
almost entirely by the canonical momentum term. However, at this
point we note that, since according to \reef{Kubo} the shear
viscosity is given by the imaginary part of the Green's function,
then the second term will not contribute as $\phi_k \phi_{-k}$ is
real. It turns out that we can discard the third term as well since
it is of $O(\omega^2)$, as we now explain.

It is an important point that in general the effective mass $M(u)$
in \reef{lineom2} is $\mathcal O(\omega^2)$ and therefore can be set
to zero in the low frequency approximation. Consider setting
$\phi(u)$ to a constant, in which case the corresponding radial
momentum \reef{CanM} automatically vanishes. The equation of motion
\reef{lineom2} must be satisfied, since a constant $\phi(u)$ simply
corresponds to a rotation and rescaling of the $x,y$ coordinates.
Therefore $M(u)$ must be $O(\omega)$, but time reversal invariance
demands that it must be proportional to $\omega^2$. We conclude that
on general grounds we must have $M(u)= O(\omega^2)$, and therefore
can be set to zero in the low frequency limit, which is taken in
calculating the shear viscosity via \reef{Kubo}. In particular, with
regard to the flux in \reef{flux}, the third term is proportional to
the mass term of the lowest order equation of motion and so is
$O(\omega^2)$. Hence this term is also irrelevant in calculating
$\eta$ in the low frequency limit. We conclude that the only
relevant piece in the flux is the canonical momentum term, and so
\be \eta = -\lim_{\omega\to 0} \frac 1{\omega}\, \mbox{Im}~
G^R_{xy,xy}(\omega,\mathbf k=0)=\lim_{u,\omega\to 0}
\frac{\Pi(u)}{i\omega \phi(u)} \,, \labell{step}.
 \ee
Here $\Pi(u)\equiv \Pi_{\lbrace\omega,\mathbf k=0\rbrace}(u)$ and
$\phi(u)\equiv \phi_{\lbrace\omega,\mathbf k=0\rbrace}(u)$. Further,
in this limit, the equation of motion \reef{lineom2} for the
canonical momentum is simply
\be
\partial_u \Pi_k(u)=0\,, \label{noflow}
\ee
\ie $\Pi_k(u)$ is independent of the radius. Therefore we are free
to evaluate the value of $\Pi(u)$ in \reef{step} at any radius and
in particular, it can be evaluated at the horizon.

Hence, the final ingredient to evaluate \reef{step} is to determine
$\omega \phi(u)$. Now to fix the fluctuations, we must impose
infalling boundary conditions on $\phi_k(u)$ at the horizon $u=u_0$.
These together with regularity at the horizon imply \cite{hong}:
\bea
\partial_u \phi(u_0,t)&=& -i \omega \left.\left(\sqrt{\frac{g_{uu}}{-g_{tt}}}
\right)\right|_{u_0}  \phi(u_0)+\mathcal O(\omega^2) \nonumber \\
\partial^2_u \phi(u_0,t)&=&-i \omega \left.\partial_u\left(\sqrt{\frac{g_{uu}}{-g_{tt}}}
\right)\right|_{u_0}  \phi(u_0)+\mathcal O(\omega^2) \nonumber \\
\partial^3_u \phi(u_0,t)&=&-i \omega \left.\partial^2_u\left(\sqrt{\frac{g_{uu}}{-g_{tt}}}
\right)\right|_{u_0}  \phi(u_0)+\mathcal O(\omega^2) \nonumber. \eea
Following \cite{hong}, keeping $\omega \phi(u)$ and $\Pi(u)$
constant in the low frequency limit, the definition of $\Pi$ implies
\be \omega \phi'(u)=\gamma (C_1(u) \phi''(u)+C_2(u) \phi'''(u))
\label{flow3}\ee
with some functions $C_1,C_2$ whose detailed form is irrelevant to
our purposes. Working perturbatively in $\gamma$,\footnote{For
certain higher curvature actions, \eg Gauss-Bonnet gravity, the
equations of motion for $\phi(u)$ are still second order in
derivatives, in which case $C_1=C_2=0$ and our conclusion follows
immediately.} one performs a split $\phi(u)=\phi_0(u)+\gamma
\phi_1(u)$. Then, to lowest order in $\gamma$, \reef{flow3} yields
the solution: $\omega \phi_0(u)=\omega \phi_0(0)$, \ie $\omega
\phi_0(u)$ is also constant in the low frequency limit. At the next
order in $\gamma$, the equation of motion for $\omega \phi_1(u)$
then also reduces to $\omega \phi_1'(u)=0$ and again with the
solution: $\omega \phi_1(u)= \omega \phi_1(0)$. We conclude that
$\omega \phi(u)=\omega \phi(0)$ to leading order in the low
frequency limit.

Hence we arrive at the result
\bea \eta &=& \lim_{\omega\to 0} \frac{\Pi(u)}{i\omega \phi(u)}=
\frac{1}{\lp^3}\left(\kappa_2(u_0)+\kappa_{4}(u_0)\right)\,.\label{Formula}
\eea
where in the second expression, we have evaluated the ratio at the
horizon $u=u_0$ and defined the quantities
 \bea \label{imp1} \kappa_2(u)=\sqrt{-\frac{g_{uu}(u)}{g_{tt}(u)}}
\left(A(u)-B(u)+\frac{F'(u)}2\right)\,, \qquad \kappa_{4}(u)
=\left(E(u)\left(\sqrt{-\frac{g_{uu}(u)}{g_{tt}(u)}}
\right)'\right)' \,.\nonumber \\
 \eea
The indices on $\kappa_\mt{i}$ indicate the number of derivatives
appearing in the corresponding terms in \reef{EfAction2}. Note that
up to now our discussion of calculating $\eta$ has been completely
general and this approach applies to any higher derivative action
involving powers of the curvature tensor (but not derivatives of the
curvature). We have verified that this approach reproduces the known
results in the literature for theories containing four-curvature
\cite{alex,buchel2} and two-curvature \cite{cool,kats} interactions.

Now we specialize the discussion to considering the action
\reef{act1}. Using the action given in equation (\ref{act1}), the
functions $A,B,E,F$ turn out to be:
\bea
\frac{A}{4\sqrt{-g}g^{uu}}&=&1+4 \c1(f_0'(u)-f_0(u)) \nonumber \\
\frac{B}{3\sqrt{-g}g^{uu}}&=&1-\frac 43\c1
\frac{f_0(u)^2-2 u^2 f_0'(u)^2}{f_0(u)} \nonumber \\
\frac{E}{\sqrt{-g}(g^{uu})^2}&=& 4 \c1 \nonumber \\
\frac{F}{\sqrt{-g}g^{uu}}&=& 16 \c1 f_0'(u) \labell{junker}
 \eea
where we have used that to lowest order in $c_{\mt i}$,
$f(u)=g(u)=f_0(u)$. With these expressions in hand, it is
straightforward to obtain the shear viscosity:\footnote{We should
point out that to obtain this result the corrections to the lowest
order background play no role other than defining the temperature
and the chemical potential. This could have been anticipated since
the corrected background only comes in at the two derivative level,
but there on general grounds the shear viscosity has the universal
form $\eta=V_3/(2l_p^3)$ where $V_3=(g_{xx})^{3/2}$ \cite{hydro}.}
 \bea
\eta &=& {\r0^3\over  2 L^3 \ell_P^3}\left( 1-8 \c1 (
f_0''-f_0')\right)\,. \labell{hatter}
 \eea
Combining this with our previous result for the entropy density
\reef{ent1}, we arrive at
 \be
{\eta \over s}={1\over 4\pi}\left(1-8 \c1 +4 (\c1+6\c2){q^2\over
\r0^6}\right)\,. \labell{etabys2}
 \ee
This agrees with the well established result when $q=0$
\cite{cool,kats}. Note that $\c3$ and $\c4$ do not appear in this
expression. We can combine the above with \reef{express0} to express
the ratio in terms of $\bar\mu=\mu/T$:
 \bea
{\eta\over s} &=&{1\over 4\pi}\left[1-8 \c1
+\frac{16\mub^2\,(\c1+6\c2)}{3\left(1+\sqrt{1+2\mub^2/3}\right)^2}\right]
 \labell{etabys3}\\
&=&{1\over 4\pi}\left[1-8 \c1+\frac{4}{3}(\c1+6 \c2)\mub^2
\left(1-\frac{1}{3}\mub^2+O(\mub^4)\right)\right] \,. \nonumber
 \eea

We note that the infalling boundary conditions are modified for the
extremal black holes and so in principle, our computation would need
to be modified in this case. Nevertheless we can consider our result
\reef{etabys2} in the extremal limit,\footnote{We can use the
leading order result found from \reef{temp0} here since $q$ only
appears in the correction term in \reef{etabys2}.} where
 \ie $q^2/\r0^6\to 2+O(c_{\mt i})$. This limit yields at $T=0$:
 \be
{\eta \over s}={1\over 4\pi}\left(1+48 \c2\right) \,.
 \labell{extremelim}
 \ee
Notice that the leading correction is now independent of $\c1$
unlike the $T\neq 0$ case.

\subsection{Conductivity and higher derivative terms}
\label{conduct}

We now turn to computing the DC conductivity in the perturbative
background corrected by the four-derivative interactions. This can
be obtained by using a Kubo formula similar to the one for the shear
viscosity. Let us define
 \be
G^R_{x,x}(\omega,\mathbf k=0)=-i \int dt d{\mathbf x}\,e^{i\omega t}
\theta(t)\,\langle [J_x(x), J_x(0)]\rangle\,.\labell{GreenA},
 \ee
where $J_\mu$ is the CFT current dual to the bulk gauge field
$A_\mu$. Then the DC conductivity is given by:
\be \sigma=-\lim_{\omega\to 0} \frac{e^2\, L_*^2}{\omega}\,
\mbox{Im}~ G^R_{x,x}(\omega,\mathbf k=0)\,.
 \labell{Kubo2}
 \ee
Here the factor $L_*^2$ appears in the prefactor so that $\sigma$
corresponds to the conductivity of the current dual to the properly
normalized potential $\tilde A_\mu$. Further, in order to interpret
the result as the `electrical' conductivity of the plasma, we
imagine coupling the CFT current to an external or auxiliary vector
field, following \cite{adam2,dilepton}. This auxiliary vector gauges
the corresponding global $U(1)$ symmetry in the CFT with a (small)
coupling $e$. Then to leading order in $e$, the effects of the
auxiliary vector are negligible and the conductivity can be
determined from the original CFT alone. The same result can also be
related to the thermal conductivity \cite{ason} which determines the
response of the heat flow to temperature gradients, \ie $T^t{}_i=
-\kappa_\mt{T}\,\partial_i T$ in the hydrodynamic
limit.\footnote{The full relativistic expression for the heat flow
also includes a contribution proportional to the pressure gradient
\cite{ason}.} The full expression for the thermal conductivity can
be written as \cite{ason}
 \be
\kappa_\mt{T}=\left(\frac{s}{n_q}+\frac{\mu}{T}\right)^2
\frac{T}{e^2}\,\sigma\,. \labell{thermcond}
 \ee

The computations are most conveniently performed within an effective
action approach, as in the previous subsection. Since the $A_t$
component of the bulk vector is nonvanishing in the background
\reef{bkgd}, the perturbations $A_x$ can couple to the shear mode
graviton, \ie metric perturbations of the form $h_{xi}$. However,
gauge invariance imposes a relation between the two sets of
perturbations which we use to integrate out the $h_{xi}$ and obtain
an action that involves only the $A_x$ fluctuation.

Starting with the action \reef{act1}, we compute the quadratic
action for the perturbations
\bea
h_{t}{}^x&=& \int \frac{d^4k}{(2\pi)^4}\, t_k(u)\,
e^{-i \omega t+i k z}\,,\nonumber \\
h_{u}{}^x&=& \int \frac{d^4k}{(2\pi)^4}\,  h_k(u)\,
e^{-i \omega t+i k z}\,,\labell{fluc4} \\
A_{x}&=& \int \frac{d^4k}{(2\pi)^4}\, a_k(u)\, e^{-i \omega t+i k
z}\,.\nonumber
 \eea
To begin, we consider only the leading order background,
\reef{lead1} and \reef{lead2}, setting $\c{i}=0$ in the action. We
would like to set the perturbation $h_u{}^x$ to zero as a gauge
choice. The corresponding component of Einstein's equations then
becomes a constraint which yields:
\be g_{xx}\, t_k'=-A'_t\, a_k\,.
 \labell{constraintx}\ee
Plugging this constraint back into the effective action along with
$h_k(u)=0$, the quadratic action takes the simple form:
\be \tilde I_{a}^{(2)}= \frac 1{2\lp^3} \int \frac{d^4k}{(2\pi)^4}
du \left(N(u) a_k'a_{-k}'+M(u) a_k a_{-k}\right)\labell{QuadAction},
\ee
where
\be N(u)=-\frac{\r0^2}{L^3} f_0(u), \qquad M(u)=\frac{L\,\omega^2}{4
u f_0(u)}-\frac{u}{L} A_t'(u)^2\,. \labell{MN} \ee
The equation of motion is solved near the horizon with the ansatz
\be a_k(u)=C f_0(u)^{\alpha} \ee
with $\alpha=\pm i \frac{\omega}{4\pi T}$ as usual. The infalling
boundary condition corresponds to choosing the minus sign. The
equation of motion for $a_k$ can be re-expressed as
\be
\partial_u j_k(u)=\frac{1}{\lp^3}M(u)\, a_k(u)
 \labell{flow}
\ee
where, as in the previous section, we have defined the radial
momentum for the effective scalar
\be j_k(u)\equiv\frac{\delta \tilde I_{a}^{(2)}}{\delta
a_{-k}'}=\frac{1}{\lp^3} N(u)\, a_k'(u)\,. \ee
The condition of regularity at the horizon $u=u_0$ corresponds to
setting \cite{hong}
\be j_k(u_0)=-i \omega \lim_{u\to u_0}\frac{N(u)}{\lp^3}
\sqrt{\frac{g_{uu}}{g_{tt}}} \,a_k(u_0)+\mathcal O(\omega^2)\,,
\labell{regularj} \ee
where we are expanding in small $\omega$ with the zero-frequency
limit of \reef{Kubo2} in mind. Next one evaluates the on-shell
action to identify the flux factor, which can be written as simply
\be 2\,{\cal F}_k = j_k(u)\,a_{-k}(u)\,. \labell{fluxj} \ee
The Green's function \reef{GreenA} is given by evaluating the
flux with the appropriate normalization at the asymptotic boundary.
The DC conductivity \reef{Kubo2} is then given by a formula
analogous to \reef{step} for the shear viscosity,
\be \sigma = \lim_{u,\omega\to 0} \frac {e^2 L_*^2}{\omega}\,
\mbox{Im}\,\left[\frac{2\,{\cal
F}_k}{a_k(u)a_{-k}(u)}\right]_{\mathbf k=0}=e^2
L_*^2\lim_{u,\omega\to 0}
\left.\frac{\mbox{Im}[j_k(u)a_{-k}(u)]}{\omega
a_k(u)a_{-k}(u)}\right|_{\mathbf k=0} \,, \labell{step2} \ee
where it is convenient not to cancel the factors of $a_{-k}(u)$ in
the final expression, as will become apparent below. The key
difference between the present case and the computation of the shear
viscosity is that neither $j_k$ nor $\omega\, a_k$ is independent of
the radial position, even in the low frequency limit. As is evident
from \reef{MN}, the effective mass $M(u)$ no longer vanishes in this
limit and so the equation of motion \reef{flow} still
produces a nontrivial flow in the radial ridection. However, if we apply
\reef{flow} in examining the radial evolution of the numerator in
\reef{step2}, we find
\be \frac{d}{du}\mbox{Im}[j_k(u) a_{-k}(u)]=\mbox{Im}\left(f_1(u)
a_{k}a_{-k}+f_2(u) j_k j_{-k}\right)=0\,. \ee
Notice that this result does not rely on the fact that we are taking
a low frequency limit. Therefore we are free to evaluate
$\mbox{Im}[j_k(u) a_{-k}(u)]$ at any radius, \eg at the horizon.
However, at the horizon, $j_k$ is constrained by the regularity
condition \reef{regularj} and so we may write
\be \sigma= \frac{e^2 L_*^2}{\lp^3}\,\kappa^A_2(u_0)
\left.\frac{\mathcal N(u_0)}{\mathcal N(0)}\right|_{\mathbf
k=0}\,,\labell{frog} \ee
where we have defined
\be \kappa^A_2(u)=-N(u)\sqrt{\frac{\,g_{uu}}{-g_{tt}}} \qquad{\rm
and}\qquad \mathcal N_k(u)=a_{k}(u)a_{-k}(u)\,. \ee
The quantity $\mathcal N(u)$ is real and so independent of $\omega$
up to $\mathcal O(\omega^2)$. This also means that to this order,
$\mathcal N(u)$ is completely regular at the horizon, since the
logarithmic divergence of $a_k(u)$ there is always accompanied by a
factor of $i\omega/T$. Therefore, in computing $\mathcal N$, we are
free to solve $a_k(u)$ imposing regularity at the horizon and
setting $\omega$ to zero, which simplifies the calculation
considerably. In the leading order background
\reef{bkgd1}--\reef{lead4}, the solution is easy to obtain:
\be a_k(u)=a_k(0)\frac{1+\frac{q^2}{2\r0^6}(2-3u)}{1
+\frac{q^2}{\r0^6}}\,. \labell{aflow}\ee
The leading order conductivity then follows:
 \bea
\sigma&=&\frac{e^2L_*^2}{2\, \lp^3}\frac{\r0}{L}\left(\frac{1-
\frac{q^2}{2\r0^6}}{1+\frac{q^2}{\r0^6}}\right)^2
\labell{conduct00}\\
&=&\frac{\pi^3e^2}{4}\frac{L^3}{\lp^3}T\frac{\left(
1+\sqrt{1+\frac{2}{3}\mub^2}\right)^3}{ \left(1+\mub^2+
\sqrt{1+\frac{2}{3}\mub^2}\right)^2}\,,
 \nonumber\eea
using our previous formulae \reef{choice}--\reef{express0}.

Extending these calculations to work at first order in the
four-derivative couplings $\c{i}$ is straightforward. One follows
the same steps as above. That is, one first computes the quadratic
effective action for $a_k,t_k,h_k$ and obtains a constraint upon
setting $h_k=0$. Substituting the constraint back into the action
and keeping terms linear in $\c{i}$ one still gets an action of the
form \reef{QuadAction}. In the subsequent steps to calculate
$\sigma$, the equation of motion for $a_k$ is technically harder to
solve and so we only present the results to leading order for small
$\mub$, \ie to order $O(\mub^2)$ or $O(q^2)$. The conductivity then
turns out to be
 \bea
\sigma &=&\frac{e^2L_*^2}{2\, \lp^3}\frac{r_0}{L}\left(1+16
\c2+\frac{q^2}{\r0^6}(-3+68 \c1+40 \c2+96(2\c3 +\c4)\right)
\labell{cond0}\\
&=& {\pi^3 e^2 L^3 T\over 2\,
\lp^3}\left(1+\frac{5}{3}\c1+16\c2-\frac{5 \mub^2}{6}[1
-\frac{1}{5}(153\c1+112\c2+192
(2\c3+\c4))]+O(\mub^4)\right)\,.\nonumber
 \eea

Now following \cite{adam2}, we examine the ratio
 \be
\frac{\sigma\, T^2}{\eta\,e^2}=1-\frac{4}{3}\mub^2-
\frac{10}{3}\c1+16\c2+\frac{8}{3}\mub^2[13\c1+
\c2+12(2\c3+\c4)]+O(\mub^4)\,. \labell{boom}
 \ee
As described in the introduction, \cite{adam2} suggested that the
simplicity of the leading result at $\mub=0$, \ie $1$, may indicate
that this result is universal. Further they argued that this leading
behaviour may then represent a universal upper bound for this ratio.
The above result certainly indicates that $\sigma T^2/(\eta e^2)
\leq 1$ even for $\mu\neq 0$. However, the leading order result is
also modified by the four-derivative couplings and so the upper
bound conjectured here may also be violated depending on the precise
values of these couplings, similar to what was found for the KSS
bound in \cite{cool,kats}.

In \cite{adam2}, it was also noted that the ratio above depends on
the relative normalization of the current and the stress tensor in
the CFT but they further observed that this relative normalization
does not appear in the ratio of the conductivity to the
susceptibility. Hence \cite{adam2} suggested that it may be more
natural for the latter ratio to obey a universal bound. To examine
how the higher derivative couplings effect this ratio, we must first
calculate the susceptibility $\Xi$:
 \be
\Xi(T)\equiv\left.\frac{\partial n_q}{\partial\mu}\right|_T\,.
 \labell{xi}
 \ee
where $n_q$ is defined in \reef{char0}. By using the result
\reef{cond0} for $\sigma$, we arrive at the following expression for
$\sigma/\Xi$:
 \be
\frac{\sigma}{\Xi}={e^2\over 2\pi T}
\left(1-\frac{11}{6}\mub^2-2\c1+16\c2
+\frac{2}{3}\mub^2\left[\frac{115}{3}\c1-32\c2+12(2\c3+\c4)
\right]+O(\mub^4)\right)\,. \labell{ratio}
 \ee
It was conjectured that ${e^2/ (2\pi T)}$ would be the lower bound
for $\sigma/\Xi$ in \cite{adam2}, at least when $\mub=0$. However,
this possibility must again be questioned in light of the fact that
the universal behaviour observed to leading order is again affected
by the four-derivative couplings.

Finally, let us turn to the thermal conductivity \reef{thermcond}.
The following interesting ratio was constructed in
\cite{ason}:\footnote{Note that the leading order result given in
\cite{ason} was $8\pi^2$ because their normalization for the gauge
kinetic term differs by a factor of 2 from that used here.}
 \be
\frac{\kappa_\mt{T}\, \mu^2}{\eta\, T}=4 \pi^2\left( 1 + \frac{2}{3}
(23 \c1 + 24 \c2)  +
 \frac{2}{9} (37 \c1 + 36 (\c2 + 4 \c3 + 2 \c4)) \mub^2 +O(\mub^4)\right)\,.
 \ee
Here we note that again the four-derivative interactions modify the
leading behaviour but, in particular, also introduce a dependence on
$\mub$, similar to what was found for the ratio $\eta/s$ in
\reef{etabys3}.

\section{Discussion}\label{discuss}

In this paper, we calculated the thermal and hydrodynamic properties
of the CFT plasma dual to a charged planar AdS black hole. These
calculations were made within the perturbative framework where the
leading Einstein-Maxwell action was extended to include a general
set of four-derivative interactions \reef{act1}. We should say that
this analysis partially overlaps with previous results and so let us
briefly summarize what was already known in the literature: In
\cite{RF2}, the authors considered the transport properties of the
charged planar AdS black hole solution with the standard
two-derivative action, including the electrical conductivity
$\sigma$. Our results agree with theirs. Born-Infeld black holes
were considered in \cite{caisun}, \ie the two-derivative action was
extended to include the combination of four-$F$ terms arising from
the expansion of the DBI action. They found that $\eta/s=1/4\pi$
which is obtained straightforwardly from \reef{etabys3} by observing
that the ratio is independent of $\c3$ and $\c4$. The effect of
adding a Gauss-Bonnet term to the gravitational action was
considered in \cite{GBF2} and $\eta/s$ was calculated in the charged
planar AdS black hole background. In \cite{GBF4}, this was extended
to include both the Gauss-Bonnet term and the two independent
four-$F$ terms. In both cases, our results are in agreement with
theirs. The work was also extended in \cite{cai8} to include a
dilaton coupling to the Gauss-Bonnet term. However, in as discussed
in \cite{beyond}, this extra coupling does not effect our results
with the present perturbative approach. The authors of \cite{noted}
had previously considered the thermodynamics with Gauss-Bonnet
gravity and the four-$F$ terms for charged AdS black holes with
flat, spherical and hyperbolic horizons. Our results for the
thermodynamic behaviour agrees with theirs for the flat case.
Finally in \cite{adam}, the effect of the $R_{abcd}F^{ab}F^{cd}$
interaction was considered on $\sigma$ when $\mu=0$. Their result is
reproduced by setting $\c1=0=\mub$ in \reef{cond0}. Hence our
comprehensive analysis agrees with all previous results where it
should.

As noted in subsection \ref{thermo}, our result for the entropy
density \reef{ent1} agrees with the result found using Wald's
formula for higher curvature theories \cite{walds}. As expected
then, the `geometric' expression for the entropy only involves $\c1$
and $\c2$ since it is only for these couplings that the
corresponding four-derivative interactions in \reef{act1} involve
the curvature tensor. It is a nontrivial result that the couplings
$\c3$ and $\c4$ still do not enter when this result is re-expressed
in terms of the temperature and chemical potential. We might note
that these two coefficients appear in \reef{temp1} and \reef{chem1}
and so a nontrivial cancelation is required for the final result in
\reef{ent1} to be independent of these parameters. It is interesting
to observe that a similar cancelation occurs for $\c2$ when the
entropy is expressed in terms of $T$ and $\mu$, so that $s$ is also
independent of this parameter. Referring back to \reef{relate1},
this means that the entropy density is the sum of two contributions
proportional to each of the central charges $a,\,c$ appearing in the
CFT
 \be
s=\frac{9\pi^2}{2} c\,T^3\,g(\mub)-\frac{5\pi^2}{2}a
\,T^3\,h(\mub)\,, \labell{curiouss}
 \ee
where $g(\mub=0)=1=h(\mub=0)$. The shear viscosity also depends only
on $\c1$ and $\c2$, as can be be inferred from \reef{etabys2} or
\reef{etabys3}. In this case, $\c2$ still appears in the result when
it is expressed in terms of $T$ and $\mu$ but again $\c3$ and $\c4$
do not appear. That is, the shear viscosity in the CFT depends on
the central charges $a,\,c$ but also the coupling of the stress
tensor to the $U(1)$ current parameterized by $\c2$.

As noted above, only $\c1$ and $\c2$ appear in corrections to the
shear viscosity. This result is in keeping with the spirit of the
recent work in \cite{rami}. This work considers generalized higher
curvature theories of gravity and attempts to relate the shear
viscosity to a `gravitational coupling' evaluated at the horizon
with a Wald-like formula \cite{walds}. While the conjectured
formulae reproduce known results for certain higher curvature
actions, it is known that these expressions fail to reproduce the
correct shear viscosity in complete generality \cite{naba,fup}.

All four of the couplings, $\c1$, $\c2$, $\c3$ and $\c4$, appear in
our expressions for the charge density \reef{char1} and conductivity
\reef{cond0}, as well as the free energy and energy densities. Since
$\c3$ and $\c4$ parameterize couplings in the four-point function of
the $U(1)$ currents, it is natural that they play a role in
correcting the properties of the CFT plasma directly related to the
corresponding charge. It is interesting to note that these two
couplings only appear in \reef{char1} and \reef{cond0} in the
combination $2\c3+\c4$. In fact, examining all of our expressions in
sections \ref{eom} and \ref{hydro}, one finds that it is only this
particular combination of $\c3$ and $\c4$ that appears everywhere.
Hence if we organized the four-$F$ interactions in the action in
terms of $(F^2)^2-2F^4$ and, say, $F^4$, then the first term would
completely decouple from the present analysis. This observation was
previously noted in \cite{noted}. We should add that this
combination seems only to be distinguished by the particular black
holes that we are considering here. For example, this combination
does not appear as the four-$F$ interaction in the low-energy
expansion of the Dirac-Born-Infeld action \cite{arkady} or in the
four-derivative extension of the $N=2$ supergravity action
\cite{mich}.

In the extremal limit \reef{extrem1}, the entropy density
\reef{ent1} becomes
 \bea
s&=&\frac{2 \pi  \r0^3}{\lp^3L^3}\left(1 -48 ( \c1+ \c2) \right)\,,
 \nonumber\\
&=&\frac{\pi^4L^3}{3\sqrt{6}\,\lp^3} \mu^3 \left(1-\c1-24\c2 \right)
 \labell{ent1x}
 \eea
This result reflects the fact that the extremal black hole still has
a finite size horizon. The interpretation in terms of the dual gauge
theory is that even at zero temperature, a finite chemical potential
will produce a deconfined `plasma' in the dual CFT. It is
interesting to note that the ratio $\eta/s$ for this extremal
plasma, given in \reef{extremelim}, only depends on $\c2$. We should
say that we expect that this `exotic' behaviour of the CFT at zero
temperature reflects our restriction of including only the metric
and a single vector in the gravitational theory. That is, in many
supergravity scenarios, the gauge kinetic terms will couple to
various scalars and from the dual CFT perspective, such couplings
reflect a nontrivial three-point function mixing two currents with
some scalar operator. In such a scenario, the area of the horizon
typically shrinks to zero size in the extremal limit \cite{scale0}
and so a finite temperature would be required to produce a
deconfined plasma.

In our general analysis, the coefficients $\c{i}$ are treated as
independent couplings. As described above, $N=2$ supergravity in
five dimensions provides a particularly interesting class of
theories, since the super-graviton multiplet also contains a $U(1)$
vector. In this case, the bosonic action for the metric and this
vector takes precisely the form given in \reef{act1} with
$\k=1/4\sqrt{3}$. However, in this case, the bulk supersymmetry is
sufficiently restrictive to constrain all of these four-derivative
couplings to be proportional to a single overall constant. Given the
result in \reef{relate1}, this means that all of these coefficients
are proportional to $(c-a)/c$ where $a$ and $c$ are the central
charges of the dual supersymmetric CFT. In this case, the vector in
the supergravity muliplet is dual to the CFT's $R$-current, which is
in the same supermultiplet as the stress tensor \cite{hm}. Hence the
previous result is in agreement that the observation that two- and
three-point functions of these two operators in the CFT (at zero
temperature and chemical potential) are parameterized entirely by
the two central charges \cite{anselmi}.

After examining the supergravity action \cite{mich} in more detail
in appendix \ref{redef}, we find that
 \be
\c2=-\frac{1}{2}\c1\simeq-\frac{1}{16}\frac{c-a}{c}\,. \labell{ugh}
 \ee
Hence for these theories, from \reef{etabys3}, the ratio $\eta/s$
becomes
 \be
{\eta\over s} ={1\over 4\pi}\left[1-8 \c1
-\frac{32\mub^2\,\c1}{3\left(1+\sqrt{1+2\mub^2/3}\right)^2}\right]
 \labell{etabys4}
 \ee
in the presence of a chemical potential. It is clear that the sign
of the third term is controlled by $\c1$ and in fact, this sign will
be the same as that appearing in the second term. Hence if $\c1$ is
positive, both of these contributions lead to a violation of the
conjectured KSS bound \cite{kss}. So in this particular class of
theories, introducing a chemical potential only makes the violation
stronger.\footnote{A similar result appears in \cite{hana}.} For example, we note that for large
$\mub$ \reef{etabys4} yields: $\eta/s \simeq 1/(4\pi)\,\left(1-24\c1
+ O(\c1/\mub^2)\right)$. We should also add that in fact it was
found that $c>a$ and hence $\c1>0$ for all of the examples of
superconformal gauge theories examined in \cite{beyond}. Hence it
appears that such violations of the KSS bound should be considered
generic rather the exception to rule. We return to this point below.

It is also interesting to examine the bounds conjectured in
\cite{adam2} for the conductivity in the case of supergravity. If we
restrict our attention to $\mu=0$, as was explicitly considered in
\cite{adam2}, \reef{boom} and \reef{ratio} reduce to
 \bea
 \left.\frac{\sigma\, T^2}{\eta\,e^2}\right|_{\mu=0}&=& 1-
\frac{10}{3}\c1+16\c2\,,
 \labell{boom2}\\
\left.\frac{\sigma}{\Xi}\right|_{\mu=0}&=&{e^2\over 2\pi T}
\left(1-2\c1+16\c2\right) \,. \labell{ratio2}
 \eea
Hence while the general expressions also depended on $\c3$ and
$\c4$, only $\c1$ and $\c2$ appear in both of these ratios when
$\mub$ vanishes. Now let us consider these results when we
substitute the supergravity result \reef{ugh}, $\c2=-\c1/2$, and
assume that $\c1$ is positive, as found in specific examples
\cite{kats,beyond} but is plausibly a general result. In this case,
the higher order corrections reduce the value of both of the ratios,
\reef{boom2} and \reef{ratio2}. Hence the first result is in
agreement with the conjecture that $\left.\sigma
T^2/(\eta\,e^2)\right|_{\mu=0}\le 1$. However, the second ratio was
conjectured to obey a lower bound
$\left.\sigma/\Xi\right|_{\mu=0}\ge e^2/(2\pi T)$ and so the
correction produces a violation of this conjectured bound.

The fact that the sign of contribution coming from the chemical
potential in \reef{etabys4} was controlled by $\c1$ is related, in
part, to the fact that only even powers of the chemical potential
everywhere in our analysis --- of course, $n_q$ has an overall
factor of $\mu$. This property arises naturally from the tensor
structure of gravitational action \reef{act1} and the particular
background that we are studying, \ie all of the relevant
interactions contain even powers of the field strength $F_{ab}$. Of
course, the bulk vector appears with an odd power in the two
Chern-Simons-like terms in \reef{act1} but these interactions did
not play a role in the present calculations, \eg the two couplings,
$\k$ and $\c5$, appear nowhere in our results. However, the two
derivative coupling $\k$ is known to play a role when rotation is
introduced \cite{moment}. Further, these couplings should play a
role if one considers the magneto-hydrodynamics of the CFT plasma,
\ie if we introduce both bulk electric and magnetic fields, as has
been recently studied with the AdS/CFT correspondence for
three-dimensional field theories \cite{magnetic}. Extending this
work to four-dimensional CFT's as considered here may be an
interesting direction for future research.

On the other hand, the behaviour noted above suggests that the
properties of sQGP studied at RHIC, with
$(\mu_\mt{B}/T)^2\lsim0.02$, should be almost unaffected by the
baryon chemical potential. Of course, one may question whether or
not this property of our holographic models carries over to the
sQGP. However, it seems that the appearance of only even powers of
$\mu$ should be a general feature emerging from the CPT invariance
of the underlying gauge theory, irrespective of whether the latter
has a holographic dual or is even conformal. For example, the
behaviour of the plasma with a net `quark' density must be identical
to that of the charge-conjugate plasma with a net density of
`anti-quarks'. Hence, one should expect that the thermal and
hydrodynamic properties of the gauge theory plasma should generally
be even functions of $\mu$.

There was a striking difference in the computations in subsections
\ref{shear} and \ref{conduct}. In particular, the shear viscosity
could be framed in terms of quantities which were independent of the
radius and so the latter could be evaluated at the horizon. Hence
the shear viscosity of the CFT could also be interpreted as the
shear viscosity associated to the stretched horizon in the membrane
paradigm \cite{membrane}. In contrast, the quantities determining
conductivity evolved nontrivially in the radial direction and so the
same connection could not be made to the membrane paradigm.

Our analysis of the conductivity contrasts with the discussion in
\cite{hong} which considered the conductivity with a vanishing
chemical potential. Note that in this case, the mixing observed
\reef{constraintx} vanishes and one may set $h_t{}^x=0$. Further
with $\mu=0$, as seen in \reef{MN}, the effective mass $M(u)$
vanishes in the low-frequency limit and so the radial evolution
\reef{flow} becomes trivial. Hence, in the absence of a chemical
potential, there is a simple relation between the conductivity in
the CFT and the universal conductivity of the stretched horizon,
$\sigma_\mt{mb} = e^2/g_5^2$ \cite{hong}:
\be \sigma_{\mt{CFT},\mu=0}=\sigma_\mt{mb}\,\sqrt{g_{zz}(u_0)}\,.
 \labell{oldsig}
 \ee
Since the conductivity is a dimensionful quantity, the factor
$\sqrt{g_{zz}(u_0)}$ appears to convert the length scale in the CFT
to the corresponding proper length at the horizon. Of course, our
conductivity \reef{conduct00} simplifies to reproduce this result in
the limit that $q$ (or $\mu$) vanishes.

It is interesting to consider the radial flow found with $\mu\ne0$
by `evaluating the conductivity' at an arbitrary radius, \ie
removing the limit $u\to0$ from \reef{step2}. This yields
\be \sigma(u)= \frac{e^2 L_*^2}{\lp^3}\kappa^A_2(u_0)
\left.\frac{\mathcal N(u_0)}{\mathcal N(u)}\right|_{\mathbf
k=0}=\sigma_\mt{CFT}\,\left(1-\frac{3}{2}u
\frac{\frac{q^2}{\r0^6}}{1+\frac{q^2}{\r0^6}} \right)^{-2}\,,
\labell{frog2} \ee
where in the second expression, we have evaluated the result for the
leading order background with \reef{aflow} and denoted the
conductivity in \reef{conduct00} as $\sigma_\mt{CFT}$. Hence the
radial evolution is such that $\sigma(u)$ decreases monotonically as
the radius varies from $u=1$ at the horizon to $u=0$ at the
asymptotic boundary. We also note that the boundary condition at
horizon is precisely $\sigma(u=1)= \sigma_{\mt{CFT},\mu=0}
=\sigma_\mt{mb}\,\r0/L$. Hence the membrane paradigm still sets the
inner boundary condition for the nontrivial radial evolution but in
general then, $\sigma_\mt{CFT}\le \sigma_\mt{mb}\,\r0/L$, where the
equality is only achieved when $\mu=0$ and there is no radial
evolution.

In general, our discussion in subsection \ref{conduct} generalizes
the discussion of \cite{hong} to include both a finite chemical
potential and the effect of higher derivative interactions. While
the simplicity of the shear viscosity computation presented there is
essentially not effected by these additional complications, there
are in fact additional simplications of the computation beyond our
presentation in subsection \ref{shear}. We will examine these
further in an upcoming paper \cite{fup}, as well giving a covariant
Wald-type formula for $\eta$.

To close, we would like to comment on a possible implication of the
`gravity as the weakest force' conjecture, \ie the recent conjecture
in \cite{weak1} that there should be a general upper bound on the
strength of gravity relative to gauge forces in quantum gravity.
This conjecture requires that there are always light `elementary
particles' with a mass-to-charge ratio smaller than the
corresponding ratio for macroscopic extremal black holes and so
allow the extremal black holes to decay. Recently, there has been
some discussion of this conjecture in the context of the AdS/CFT
correspondence and the implications for the spectrum of the CFT
\cite{sean,simeon} --- see also footnote 4 in \cite{adam2}. The
relation which we would like to draw relies on the further corollary
that higher derivative corrections should reduce the mass-to-charge
ratio of extremal black holes in a consistent theory of the quantum
gravity \cite{weak1,weak2}. In the present context, this suggests
computing the ratio of the energy density to charge density
$\rho_E/n_q$ in the extremal limit \reef{extrem1}
 \be
\frac{\rho_E}{n_q}=\left(\frac{\rho_E}{n_q}\right)_0 \left(1-
\frac{47\c1+48 \c2}{3}\right)\,. \labell{xxtrem}
 \ee
Here, the quantity $\displaystyle
\left(\frac{\rho_E}{n_q}\right)_0=\frac{3\sqrt{3}}{2\sqrt{2}}\frac{\r0}{\pi
L^2}$ is the leading order `classical' result. Hence the weak
gravity conjecture would impose the constraint $47\c1+48\c2 >0$.
However, if we again consider the specific case of supergravity with
$\c2=-\c1/2$, this constraint becomes simply $\c1>0$.

While the above analysis is suggestive, it misses the intent of the
discussion in \cite{weak1,weak2} which was phrased in terms of
extremal black holes in asymptotically flat space. Their natural
assumption was that the `quantum' corrections coming from higher
derivative terms in the gravitational action should decrease the
mass-to-charge ratio but also do so in a way that the decrease
becomes more pronounced for smaller (extremal) black holes. Hence a
proper comparison requires repeating the analysis of section
\ref{solut} for charged AdS black holes with spherical horizons.
This is a straightforward exercise and the final result replacing
\reef{xxtrem} is
 \be
\frac{\rho_E}{n_q}=\left(\frac{\rho_E}{n_q}\right)_0 \left(1-
\c1\,f(\r0/L)\right)\quad{\rm with}\
f(\r0/L)=\frac{7+34\frac{\r0^2}{L^2}+107\frac{\r0^4}{L^4}
+138\frac{\r0^6}{L^6}}{ 3\frac{\r0^2}{L^2}
\left(1+2\frac{\r0^2}{L^2}\right)
\left(2+3\frac{\r0^2}{L^2}\right)}\,. \labell{xxtrem2}
 \ee
Above $\r0$ is the position of the horizon in coordinates where the
area of the spherical horizon is $2\pi^2\r0^3$. In fact, a full
analysis yields a ratio which depends on all of the $\c{i}$ but in
presenting \reef{xxtrem2}, we have focussed on the supergravity case
where all of these dimensionless parameters are proportional to
$\c1$. One easily verifies that in the limit of large black holes,
\ie $\r0/L\gg1$, this result reduces to that for the planar black
holes given in \reef{xxtrem} when $\c2=-\c1/2$. It is also evident
that $f(\r0/L)$ is positive for all values of $\r0/L$ and hence the
higher derivative corrections always reduce the mass-to-charge ratio
of these extremal black holes as long as $\c1>0$. As desired, this
effect is also largest for small black holes because of the factor
of $\r0^2/L^2$ in the denominator of $f(\r0/L)$. However, it is
interesting that $f(\r0/L)$ does not decrease monotonically as
$\r0/L$ grows. Instead the function exhibits a local minimum near
$\r0\sim L$, which seems to be an effect of the asymptotic AdS
geometry.

In certain cases, it is well understood that there are
supersymmetric bound states of giant gravitons \cite{giant} carrying
the same charges as these charged black holes with spherical
horizons. Further that there is a gap in the spectrum between the
extremal black hole and these bound states \cite{giant2}. Hence one
can understand the details of realizing the `gravity as the weakest
force' conjecture within this framework. However, our interest in
these issues comes rather from the general constraint $\c1>0$. In
particular with \reef{relate1}, we can infer that the weak gravity
conjecture requires an inequality for the central charges of any
four-dimensional conformal field theory with a gravitational dual,
namely $c>a$. Of course, this is precisely the inequality which was
observed for the broad class of superconformal gauge theories
examined in \cite{beyond}. Hence it seems there may be a deep
connection between this provisional observation for superconformal
gauge theories and the consistency of their holographic duals as a
theories of quantum gravity.

\acknowledgments It is a pleasure to thank Alex Buchel, Sera
Cremonini, Ben Freivogel, Sean Hartnoll and Uli Heinz for useful
correspondence and conversations. Research at Perimeter Institute is
supported by the Government of Canada through Industry Canada and by
the Province of Ontario through the Ministry of Research \&
Innovation. RCM also acknowledges support from an NSERC Discovery
grant and funding from the Canadian Institute for Advanced Research.
MFP is supported by the Portuguese Fundacao para a Ciencia e
Tecnologia, grant SFRH/BD/23438/2005.

\appendix

\section{Field Redefinitions and the four-derivative action}\label{redef}

In this appendix, we will demonstrate that our action \reef{act1}
contains the most general four-derivative interactions involving a
single Maxwell field. We should add that a similar analysis appeared
in \cite{jiml} but they began with a more restricted starting point.
First, we consider the leading order two-derivative action
 \beq
 I_\mt{2}=\frac{1}{2\lp^3}\int d^5x
\sqrt{-g}\left[\,\frac{12}{L^2} + R
-\frac{1}{4}F^2+\frac{\k_1}{3}\veps^{abcde}A_aF_{bc}F_{de}\right]\,.
 \labell{act02}
 \eeq
Next the most general four-derivative action for gravity coupled to
a single $U(1)$ vector takes the form:
 \beqa
 I_\mt{4}&=&\frac{L^2}{2\lp^3}\int d^5x
\sqrt{-g}\left[\,\al_1 R^2+\al_2 R_{ab}R^{ab}+\al_3 R_{abcd}R^{abcd}
 \right.
 \labell{act04}\\
&&\qquad+ \b_1 R F^2 + \b_2 R^{ab}F_{ac}F_b{}^c+ \b_3
R_{abcd}F^{ab}F^{cd}+ \b_4 R_{abcd}F^{ac}F^{bd}
 \nonumber\\
&&\qquad+ \del_1 \left(F^2\right)^2+ \del_2\,F^4 + \del_3
\na^aF_{ab} \na^cF_c{}^b +
\del_4 \na_aF_{bc} \na^aF^{bc} \nonumber\\
&&\qquad + \del_5 \na_aF_{bc} \na^bF^{ac}+ \del_6 \na^2 F_{ab}\,
F^{ab} + \del_7 \na_a\na^bF_{bc}\, F^{ac} + \del_8
\na^b\na_aF_{bc}\,
F^{ac} \nonumber\\
 &&\left.\qquad+ \veps^{abcde}\left( F_{ab} \left( \ga_1
F^{cd}\na^f\!F_{fe} + \ga_2 F_{cf} \na^f\! F_{de} + \ga_3 F_{cf}
\na_d F_e{}^f\right) +\k_2 \, A_a R_{bcfg} R_{de}{}^{fg}\right)
\right]\,,
 \nonumber
 \eeqa
where, as in the main text, we use $F^2=F_{ab}F^{ab}$ and
$F^4=F^a{}_bF^b{}_cF^c{}_dF^d{}_a$. Here all of the coefficients,
$\al_\mt{i},\ \b_\mt{i},\ \del_\mt{i}, \ga_\mt{i}$ and $\k_2$ are
dimensionless constants, that we expect are generically very small.
Many of the four-derivative terms above can be eliminated by simply
integrating by parts. For example,
 \beqa
&&\int d^5x \sqrt{-g}\, \na^aF_{ab} \na^cF_c{}^b
 \labell{examp1}\\
&&\qquad=\int d^5x \sqrt{-g} \left(\na_aF_{bc} \na^bF^{ac} -
R^{ab}F_{ac}F_b{}^c + R_{abcd}F^{ac}F^{bd}\right)\,.
 \nonumber
 \eeqa
Using integration by parts, as well as the identities
$\na_{[a}F_{bc]}=0=R_{[abc]d}$, one can eliminate $\b_4$,
$\del_{4,5,6,7,8}$ and $\ga_{2,3}$. In this way, the general
four-derivative action can be reduced to
 \beqa
 I_\mt{4}&=&\frac{L^2}{2\lp^3}\int d^5x
\sqrt{-g}\left[\,\al_1 R^2+\al_2 R_{ab}R^{ab}+\al_3 R_{abcd}R^{abcd}
 \right.
 \labell{act14}\\
&&\qquad+ \b_1 R F^2 + \b_2 R^{ab}F_{ac}F_b{}^c+ \b_3
R_{abcd}F^{ab}F^{cd}+ \del_1 \left(F^2\right)^2+ \del_2 F^4
 \nonumber\\
&&\left.\qquad + \del_3 \na^aF_{ab} \na^cF_c{}^b +\ga_1
\veps^{abcde} F_{ab} F_{cd}\na^fF_{fe} +\k_2 \veps^{abcde} A_a
R_{bcfg} R_{de}{}^{fg} \right]\,.
 \nonumber
 \eeqa
Now consider making field redefinitions: $g_{ab}\rightarrow
g_{ab}+\delta g_{ab}$ and $A_a\rightarrow A_a+\delta A_a$. The most
general field redefinition involving two-derivative contributions
can be written
 \beqa
\delta g_{ab} &=& \mu_1L^2\,R_{ab}+\mu_2L^2\,F_{ac}F_b{}^c+\left(
\mu_3 L^2 R +\mu_4L^2 F^2 +\mu_5\right) g_{ab}\,,
 \labell{newer1}\\
\delta A_a &=& \l_1 A_a + \l_2 \na^bF_{ba} + \l_3 \veps_{abcde}
F^{bc} F^{de} \,.
 \nonumber
 \eeqa
Note that $\mu_5$ and $\l_1$ give a (constant) rescalings of the
metric and vector, respectively, which will prove useful in the
following. We also note that other than the rescaling the field
redefinition of the vector involves covariant terms so that the
modified action remains invariant with standard gauge
transformations. In general, we might note that, aside from the two
rescalings, the field redefinitions \reef{newer1} contain six
two-derivative terms while the four-derivative action contains
eleven interactions. Hence, on general grounds, we expect that we
will be left with five independent terms in our higher-order action.
With the field redefinitions \reef{newer1}, the leading change in
the action comes from the variation of \reef{act02}
 \beqa
\delta I_\mt{2}&=&\frac{1}{2\lp^3}\int d^5x \sqrt{-g}
\left\lbrace\left[
\left(\frac{6}{L^2}+\frac{1}{2}R-\frac{1}{8}F^2\right)g^{ab}
-R^{ab}+\frac{1}{2}F^{ac}F^b{}_c\right]\,\delta g_{ab}\right.
 \nonumber\\
&&\left.\vphantom{\frac{1}{2}}\qquad\qquad\qquad\qquad\qquad
+\left(\na_b F^{ba}+\k_1 \veps^{abcde}F_{bc}F_{de}\right)\del
A_a\right\rbrace\,.
 \labell{change2}
 \eeqa
Of course, one should note that we have integrated by parts to
produce the expressions in \reef{change2}. Now we will divide this
variation into two parts, examining separately the contributions to
the four- and two-derivative actions. Beginning with the former, one
finds:
 \beqa
 \del I_\mt{4}&=&\frac{L^2}{2\lp^3}\int d^5x
\sqrt{-g}\left[\,(\frac{\mu_1}{2}+\frac{3}{2}\mu_3) R^2-\mu_1
R_{ab}R^{ab}+
\left(-\frac{\mu_1}{8}+\frac{\mu_2}{2}-\frac{\mu_3}{8}+\frac{3}{2}
\mu_4\right) R F^2
 \right.
 \labell{dact14}\\
&&\qquad + \left(\frac{\mu_1}{2}-\mu_2\right) R^{ab}F_{ac}F_b{}^c+
\left(-\frac{\mu_2}{8}-\frac{\mu_4}{8}+8\k_1\l_3 \right)
\left(F^2\right)^2
 \nonumber\\
&&\left.\qquad+\left(\frac{\mu_2}{2}-16\k_1\l_3 \right) F^4 + \l_2
\na^aF_{ab} \na^cF_c{}^b + \left(\k_1\l_2+\l_3\right) \veps^{abcde}
F_{ab} F^{cd}\na^fF_{fe}\right]\,.
 \nonumber
 \eeqa
An obvious choice of the field redefinition parameters is then
 \beqa
&&\mu_1=\al_2,\ \mu_2=\b_2+\al_2/2,\ \mu_3=-(2\al_1+\al_2)/3,
 \nonumber\\
&&\mu_4=-\left(24\b_1-12\b_2+2\al_1-11\al_2\right)/36,\
\l_2=-\del_3,\ \l_3=-\ga_1+\k_1\del_3. \labell{value1}
 \eeqa
These choices eliminate six of the four-derivative interactions
leaving
 \beqa
 I_\mt{4}&=&\frac{L^2}{2\lp^3}\int d^5x
\sqrt{-g}\left[\,\al_3 R_{abcd}R^{abcd}+ \b_3 R_{abcd}F^{ab}F^{cd}
 \right.
 \labell{act24}\\
&&\left.\qquad+ \tilde\del_1 \left(F^2\right)^2+ \tilde\del_2 F^4
+\k_2 \veps^{abcde} A_a R_{bcfg} R_{de}{}^{fg} \right]\,,
 \nonumber
 \eeqa
where
 \beqa
&&\tilde\del_1=\del_1+\frac{1}{288}\left(2\al_1-29\al_2\right)
+\frac{1}{12}\left(\b_1-\b_2\right)-8\k_1\ga_1+8\k_1^2\del_3\,,
\nonumber\\
&&\tilde\del_2=\del_2+\frac{\al_2}{4}+\frac{\b_2}{2}
+16\k_1\ga_1-16\k_1^2\del_3\,.
 \labell{shift1}
 \eeqa

One may try to be more clever with the field redefinitions in the
present case. In particular, as seen in \cite{noted} (or our
calculations in sections 2 and 3), when the background only contains
a radial electric field, the four-$F$ terms only couple to the
graviton and gauge field equations for motion with the combination
$2\del_1+\del_2$. Hence might try to make a field redefinition that
sets this combination to zero. If we go back to \reef{dact14}, we
find that the field redefinitions yield the following variation of
this linear combination:
 \beq
\Delta\left[2\del_1+\del_2\right]=\frac{1}{4}\left(\mu_2-\mu_4\right)\,.
 \labell{shift2}
 \eeq
Hence we can indeed arrange to set this combination of couplings to
zero. However, the result will be that we would not be able to
eliminate all of the $R^2$ and $R F^2$ interactions involving the
Ricci tensor and Ricci scalar. Hence such a field redefinition will
simply replace the complications of accounting for the four-$F$
interactions with those of accounting for another set of
interactions. With respect to the dual CFT, these two four-$F$
interactions define two characteristic parameters that would appear
in the four-point function of the dual current. So it seems to
natural not to field redefine them away, as the four-point function
must be invariant and the previous parameters would just appear from
exchanges between bulk three-point interactions rather than being
manifest in a four-point contact interaction. However, given this
discussion, we could arrange the four-$F$ terms in \reef{act24} as
 \beq
 \frac{L^2}{2\lp^3}\int d^5x
\sqrt{-g}\left[\frac{1}{2}\left(2\tilde\del_1+\tilde\del_2\right)
\left(F^2\right)^2+ \tilde\del_2 \left(F^4
-\frac{1}{2}\left(F^2\right)^2\right) \right]\,,
 \labell{try1}
 \eeq
in which case, the second term should not contribute in the present
calculations.

At this point, we have not yet commented on fixing the values of
$\mu_5$ and $\lambda_1$. In this regard, let us consider the effect
of the field redefinitions on varying the two-derivative action
\reef{act02}:
 \beqa
\del I_\mt{2}&=&\frac{1}{2\lp^3}\int d^5x
\sqrt{-g}\left[\,\frac{12}{L^2}\frac{5}{2}\mu_5 + R
\left(6\mu_1+30\mu_3+\frac{3}{2}\mu_5\right)
 \right.\labell{dact02}\\
&&\qquad\qquad\qquad\left.
 +F^2\left(6\mu_2+30\mu_4-\frac{\mu_5}{8}-\frac{\l_1}{2}\right)+\k_1\l_1
\veps^{abcde}A_aF_{bc}F_{de}\right]\,. \nonumber
 \eeqa
Hence the convenient choice which we make in fixing these parameters
is to set
 \beq
\mu_5=-4\left(\mu_1+5\mu_3\right)\,, \qquad
\l_1=\mu_1+12\mu_2+5\mu_3+60\mu_4\,.
 \labell{value2}
 \eeq
This choice leaves the Planck scale fixed, as well as the
coefficient for the vector kinetic term. Hence after the field
redefinitions, the two-derivative action \reef{act02} becomes
 \beq
 I_\mt{2}=\frac{1}{2\lp^3}\int d^5x
\sqrt{-g}\left[\,\frac{12}{\tilde L^2} + R
-\frac{1}{4}F^2+\frac{\tilde
\k_1}{3}\veps^{abcde}A_aF_{bc}F_{de}\right]
 \labell{act12}
 \eeq
where
 \beqa
\frac{1}{\tilde L^2}&=&\frac{1}{L^2}\left(1-10 \left(\mu_1
+5\mu_3\right)\right)
 \labell{shift3}\\
\tilde\k_1&=&\k_1\left(1+3\left(\mu_1 +12\mu_2 +5\mu_3 +60\mu_4
\right)\right) \,.
 \nonumber
 \eeqa

Next we would like to apply this analysis to compare the
four-derivative supergravity action given in \cite{mich} to the
effective action \reef{act1} studied in the present
paper.\footnote{Note that this comparison requires care since
\cite{mich} adopts the supergravity conventions of \cite{sugra},
with,  \eg the mostly minus convention for the signature of the
metric. We thank Sera Cremonini for her detailed explanation of
these conventions.} Of course, the leading two-derivative terms in
the supergravity action take precisely the form given in
\reef{act02} with $\k_1=1/4\sqrt{3}\left(1-32\c1\right)$ -- here and
in the following, we present the results in terms of
$\c1=(c-a)/(8c)$, as in \reef{relate1}. The four-derivative
supergravity action can certainly be described in terms of our
general action \reef{act04} where the dimensionless coefficients are
all assigned specific values proportional to $\c1$. However, the
supergravity action is naturally described in terms of the Weyl
tensor and so to make this matching, we must first re-express the
following:
 \beqa
C_{abcd}\,C^{abcd}&=&R_{abcd}R^{abcd}-\frac{4}{3}R_{ab}R^{ab}
+\frac{1}{6}R^2\,,
 \labell{wyle}\\
C_{abcd}F^{ab}F^{cd}&=& R_{abcd}F^{ab}F^{cd}
-\frac{4}{3}R^{ab}F_{ac}F_b{}^c +\frac{1}{6}R F^2\,,
 \nonumber
 \eeqa
which applies for the five-dimensional Weyl tensor. Now as described
above, integration by parts can be used to eliminate many terms in
the general action \reef{act04}, reducing it to the form given in
\reef{act14}. Putting the supergravity action in this particular
form yields:
 \beqa
\alpha_1&=&\frac{1}{6}\c1\,, \qquad \quad \ \
\alpha_2=-\frac{4}{3}\c1\,,
\qquad \alpha_3=\c1\,, \nonumber\\
\beta_1&=&\frac{1}{4}\c1\,, \qquad \quad \ \
\beta_2=-\frac{4}{3}\c1\,, \qquad \ \ \
\beta_3=-\frac{1}{2}\c1\,, \labell{lotsa1}\\
\delta_1&=&-\frac{41}{288}\c1\,, \qquad \delta_2=\frac{5}{8}\c1\,,
\qquad \ \ \ \delta_3=2\c1\,,\nonumber\\
\gamma_1&=&\frac{5}{8\sqrt{3}}\c1\,, \qquad\
\kappa_2=\frac{1}{2\sqrt{3}}\c1\,.
 \eeqa
Now applying the field redefinitions \reef{newer1} with the
parameters fixed as in \reef{value1}, we are able to further reduce
the four-derivative action to the canonical form \reef{act24}. While
these field redefinitions leave the values of $\alpha_3$, $\beta_3$
and $\kappa_2$ unchanged from those given above, \reef{shift1}
yields
 \beq
\tilde\delta_1=\frac{1}{24}\c1\,,\qquad
\tilde\delta_2=-\frac{5}{24}\c1\,. \labell{lotsa2}
 \eeq
As a final step, we give the Einstein and Maxwell kinetic terms
their standard normalization with \reef{value2} yielding
 \beq
\frac{1}{\tilde L^2}=\frac{1}{L^2}\left(1-\frac{10}{3}\c1
\right)\,,\qquad
\tilde\k_1=\k_1\left(1-256\c1\right)=\frac{1}{4\sqrt{3}}\left(1-288\c1\right)
\,.
 \labell{shift4}
 \eeq
Hence, after all of these manipulations, the supergravity action
takes the form given in \reef{act1} with
 \beq
\c2=-\frac{1}{2}\c1\,, \quad \c3=\frac{1}{24}\c1\,, \quad
\c4=-\frac{5}{24}\c1\,, \quad \c5=\frac{1}{2\sqrt{3}}\c1\,,
 \labell{finlotsa}
 \eeq
as well as $\k=1/4\sqrt{3}\left(1-288\c1\right)$ and
$\c1=(c-a)/(8c)$, as in \reef{relate1}. As discussed, this latter
combination of the central charges in the (supersymmetric) CFT is
seen to explicitly fix the coefficients all of these four-derivative
corrections to the supergravity action.

\end{document}